# Rationalizing the influence of tunable energy levels on quantum efficiency to design optimal non-fullerene acceptor-based ternary organic solar cells

Safakath Karuthedath,[1,7+] Sri H. K. Paleti,[1,7] Anirudh Sharma,[1] Hang Yin,[2] Catherine S. P. De Castro,[1] Si Chen,[1] Han Xu,[1] Nisreen Alshehri,[1,3] Nicolas Ramos,[4] Jafar I. Khan,[1] Jaime Martin,[4,5] Gang Li,[2] Frédéric Laquai,[1] Derya Baran[1,*] and Julien Gorenflot[1,6,**]

[1]King Abdullah University of Science and Technology (KAUST), KAUST Solar Center (KSC), Physical Sciences and Engineering Division (PSE), Material Science and Engineering Program (MSE), Thuwal 23955-6900, Kingdom of Saudi Arabia

[2]Department of Electronic and Information Engineering, The Hong Kong Polytechnic University, Hung Hom, Kowloon, Hong Kong SAR, P. R. China

[3]Physics and Astronomy Department, College of Sciences, King Saud University, Riyadh 12372, Kingdom of Saudi Arabia

[4]POLYMAT, University of the Basque Country UPV/EHU Av. de Tolosa 72, 20018, San Sebastián, Spain

[5]Ikerbasque Basque Foundation for Science Bilbao 481013, Spain

[6]Lead contact

[7]These authors contributed equally

[+] Present address: Institute of Materials Research, Tsinghua Shenzhen International Graduate School, Tsinghua University, Shenzhen, 518055 China

* Correspondence: derya.baran@kaust.edu.sa

** Correspondence: julien.gorenflot@kaust.edu.sa

## Summary

Non-fullerene acceptor (NFA)-based ternary bulk heterojunction solar cells (TSC) are the most efficient organic solar cells (OSCs) today due to their broader absorption and quantum efficiencies (QE) often surpassing those of corresponding binary blends. We study how the energetics driving charge transfer at the electron donor:electron acceptor (D/A) interfaces impact the QE in blends of PBDB-T-2F donor with several pairs of lower bandgap NFAs. As in binary blends, the ionization energy offset between donor and acceptor ($\Delta IE$) controls the QE and maximizes for $\Delta IE > 0.5$ eV. However, $\Delta IE$ is not controlled by the individual NFAs $IE$s but by their average, weighted for their blending ratio. Using this property, we improved the QE of a PBDB-T-2F:IEICO binary blend that had an insufficient $\Delta IE$ for charge generation by adding a deep $IE$ third component: IT-4F. Combining two NFAs enables to optimize the D/A energy alignment and cells' QE without molecular engineering.

## Introduction

The power conversion efficiency (PCE) of single-junction organic solar cells (OSC) has been blooming in the latest years, now having crossed 19%, with 20% expected soon,[1] driven by the development of novel non-fullerene acceptors (NFA) and improved device optimization.[1b, 2] As of today, the most efficient single-junction OSCs are ternary organic solar cells (TSC), combining one electron donor and two NFAs in the photoactive layer.[3] TSC offer better performances[4] and stability[4] than the corresponding binary cells[5], and are easier to upscale than multi-junction solar cells.[6] Looking into the near future and the development of tandem and multijunction OSC, this 3rd component offers an additional lever to adjust the photovoltaic parameters in the subcells, as was recently demonstrated.[3b, 7]

The primary reason for the TSCs' higher performances is the improved photocurrent due to their broader absorption and, often, to an enhanced quantum efficiency of the photocurrent generation.[3b, 7b, 8] This higher quantum efficiency has been attributed to enhanced charge generation[3b] and charge carrier mobilities[7b, 9] and reduced recombination.[3b, 4a] In addition, a third component with properly selected energy levels constitutes an additional lever to optimize the driving force for photocurrent generation and minimize energy losses.

In TSC, the combination of the three components' energy levels has mainly been studied in the frame of energy losses and cells' open-circuit voltage ($V_{OC}$). Upon changing the blend ratio, $V_{OC}$ has been found to evolve linearly between the $V_{OC}$ of the binary cells corresponding to ratios 1:0:1 and 1:1:0.[10]

Additionally, when a pair of donor[11] (resp. acceptor[12]) molecules form perfect co-crystals, several studies have shown that the ionization energy (IE) of the mixture can be continuously tuned between the IE of the two components by changing their blending ratio. Zhang et al. have used this strategy to deepen the IE of an acceptor phase originally composed of ZITI-N by the addition of ZITI-S, to the point where its offset with the donor IE became sufficient to drive efficient charge generation.[12a] Here, we studied pairs of more dissimilar acceptors blended with a single donor (PBDB-T-2F) to understand the interplay of their energy levels. We rationalize the effect of those energetics on photocurrent generation in the frame of our recent findings on how the offset $\Delta IE$ between the ionization energy of a donor and acceptor molecules controls photocurrent generation efficiency in binary solar cells.[13]

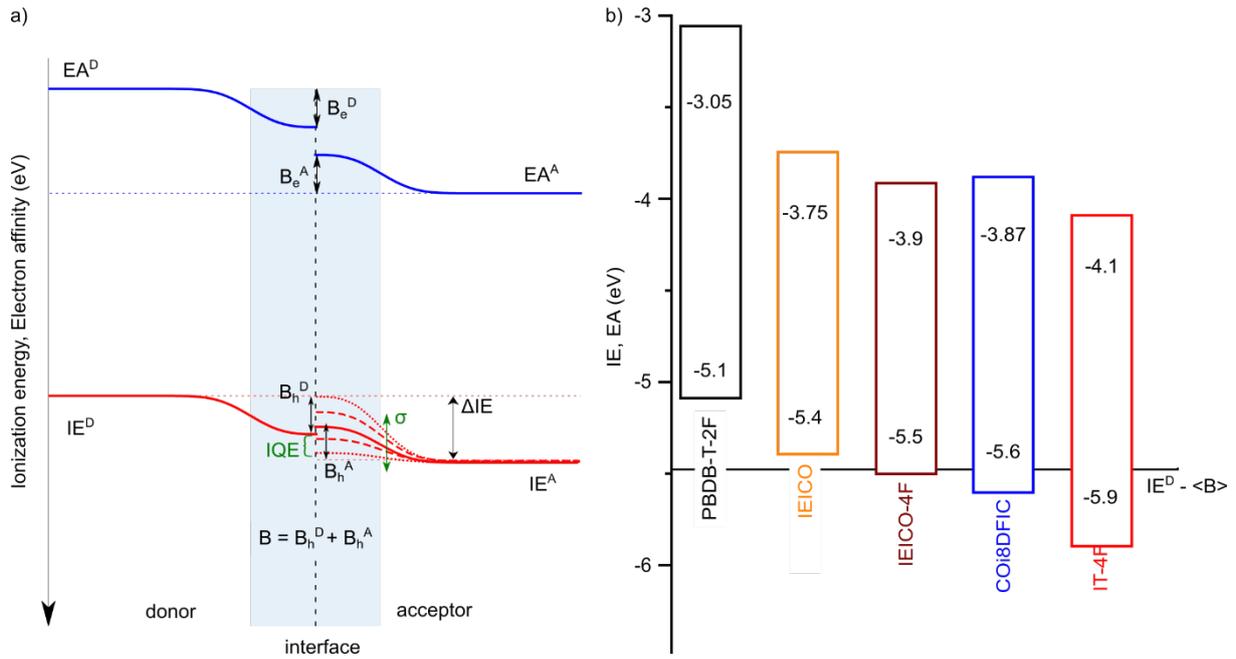

*Figure 1: In binary donor:low bandgap NFA blends, ionization energy (IE) offset and ionization energy bending at the donor:acceptor interface control the maximum reachable solar cells internal quantum efficiency. (a) schematic model of the ionization energy (IE) and electron affinity (EA) in the vicinity of a donor:acceptor interface. B represents the energy level bending, the superscripts A and D stand for acceptor and donor, respectively, and the subscripts e and h for electron and hole, respectively. Considering a Gaussian distribution of interface bending of variance $\sigma^2$ and average <B>, the maximal IQE is given by the fraction of these interfaces for which the IE in the acceptor side is deeper than in the donor side, thus enabling the hole transfer. Note that disorder is present in both the donor and the acceptor sides but is represented only on the acceptor side for simplicity. (b) Energy levels of the materials used in the study. IE and EA were measured using ultraviolet photoelectron spectroscopy (UPS) and low energy inverse photoelectron spectroscopy (LE-IPES), respectively, see Figure S2. The horizontal line at -5.49 eV represents the $IE^A$ level for which $\Delta IE$ is equal to the average bending <B> previously reported for PBDB-T-2F:NFA blends.[13]*

In binary blends, we found that energetics control not only the photovoltage, but also the photocurrent.[13-14] Specifically, an ionization energy offset ($\Delta IE$) of no less than 0.45 – 0.5 eV between the donor and the NFA can drive a quantitative hole transfer from the exciton in the NFA to the donor.[13, 15] As shown in Figure 1a, this offset ensures that the electrostatic bending of the energy levels at the donor-acceptor interface does not bring the IE of the donor below that of the acceptor at the interface, which would prevent hole transfer.[13, 15] Or equivalently that the interface charge transfer state energy remains below that of the NFA singlet excitons, thus allowing for the hole transfer to occur. Note that the singlet exciton is almost always located in the NFA even when the donor absorbs light. Indeed, by design, the NFA in modern cells has a lower bandgap than the electron donor, resulting in Förster resonance energy transfer (FRET) from the donor to the acceptor which outcompetes electron transfer.[7b, 13, 16] Note also that the interface energy level bending,[13, 15] while being a limit for the charge transfer, tends to push the charges away from the interface after the hole transfer has occurred, resulting in barrierless charge separation.[17] Finally, we note that the necessity of having $\Delta IE$ larger than the energy level bending is not an energy loss, as the interfacial state energy, and hence $V_{OC}$, is raised by the same quantity $<B>$ compared to what it would be if considering only the pristine NFA and donor energy levels. Such an understanding determines the selection of the materials that can be combined for efficient charge generation and is thus also crucial for TSC. We note that Li et al. explained the D/A interface energetics alternatively by a shift of the materials work function (measured in bilayers), due to interface dipole, with the same final consequences on solar cells operation.[18] However, the bending of the $IE$ is well documented and unlikely to be absent.[15, 17b, 19] Potentially both effects participate in the interface energy landscape.[20] For the sake of simplicity here however, we will group both effects within the energy levels bending term $B$.

In this work, we selected the common donor copolymer PBDB-T-2F (PM6) as a wide-bandgap electron donor. We used three different NFAs as the majority acceptor component (hereafter referred as NFA-1) in the 1:0.9:0.1 weight ratio D:NFA-1:NFA-2 blends which enabled us to model three different energetic scenarios (see Figure 1b): i- IEICO has a $\Delta IE$ (with PBDB-T-2F) lower than the average energy level bending $<B>$ = 0.39 eV reported for PBDB-T-2F:NFA blends.[13] ii- IEICO-4F has a $\Delta IE$ (with PBDB-T-2F) comparable to $<B>$. iii- Finally, IT-4F offers a $\Delta IE$ larger than $<B>$. As a minority NFA component (hereafter referred as NFA-2), we used COi8DFIC, which offers a $\Delta IE$ with PBDB-T-2F slightly larger than the reported value of $<B>$. Previously, we reported that in binary PBDB-T-2F:IEICO and PBDB-T-2F:IEICO-4F solar cells,[4a] the photocurrent generation is low since the IE offset between the donor and acceptor is too small to drive charge generation efficiently.[13] Here, COi8DFIC was chosen as the second acceptor due to its IE larger than those of IEICO and IEICO-4F. The chemical structures can be found in Scheme S1 of the Supporting Information.

We investigated the photophysical processes in TSCs with different IE offsets using steady-state and transient spectroscopy techniques. Similar to the binaries situation,[13] we find that FRET from the donor to the NFAs outcompetes electron transfer, leading to an $\Delta IE$-controlled charge generation (hole transfer). We find that the hole transfer from NFA-1 and COi8DFIC to PBDB-T-2F behaves as a single channel. As such, the hole transfer efficiency from COi8DFIC to PBDB-T-2F depends on the associated NFA-1. We then investigated what offset controls this unique channel. By varying the NFA-1:COi8DFIC ratio, we observed that the offset between the *IE* of the donor and the weighted average *IE* of the acceptors ($\Delta IE_{weighted}$) controls the charge generation process. This implies that even low *IE* offset D/A combinations can contribute to charge generation in ternary solar cells – unlike in binary devices – provided they are associated with a second NFA offering a larger offset.

## Results and Discussion

### 1. Device performance and steady-state optical properties

Our previous observations gives us the following expectations for the selected binary blends:[13] i- for PBDB-T-2F:IEICO-4F, $\Delta IE$ is slightly lower than the reported average $<B>$ of the Gaussian distribution of interface energy level bendings,[13] meaning that a bit less than half of the interface sites have a donor hole energy $IE_{interface}^{D}$ deeper than that of a hole in the acceptor $IE_{interface}^{A}$ (see Figure 1a) and thus favorable to hole transfer, leading to an expected IQE slightly below 50% if no additional losses appear. ii- For PBDB-T-2F:IEICO, where $\Delta IE$ is lower than $<B>$ - $\sigma$, considering the previously reported standard deviation $\sigma$ of the Gaussian distribution of energy level bending, $\sigma$ = 60 meV for PBDB-T-2F:NFA cells,[13] only a limited fraction of $IE_{interface}^{D}$ will be deeper than $IE_{interface}^{A}$ leading to a low IQE. iii- On the other hand, for PBDB-T-2F:IT-4F, where $\Delta IE$ is larger than $<B> + 2\sigma$, we expect a large IQE. iv- We finally expect an IQE slightly over 50% for PBDB-T-2F:COi8DFIC where $\Delta IE$ is slightly larger than the reported $<B>$. As a reminder, we do not expect charge separation to decrease the quantum efficiency further as the energy level bending tends to increase the energy of the interfacial state, thus compensating its binding energy and making its separation favorable, or even sometimes, reportedly barrierless.[17a] Additionally, electron affinity was found to be irrelevant, as the design of modern NFA-based solar cell (larger bandgap donor and lower bandgap acceptor) is such that excitons generated in the donor typically undergo energy transfer to the NFA prior to reaching the interface (we reported a Förster Resonance Energy Transfer – FRET – radii ranging from 2.7 to 3.3 nm for the NFA-1 studied here).[13] As a result, electron transfer – potentially controlled by the electron affinity – is limited to donor excitons generated in the direct vicinity of the D/A interface.

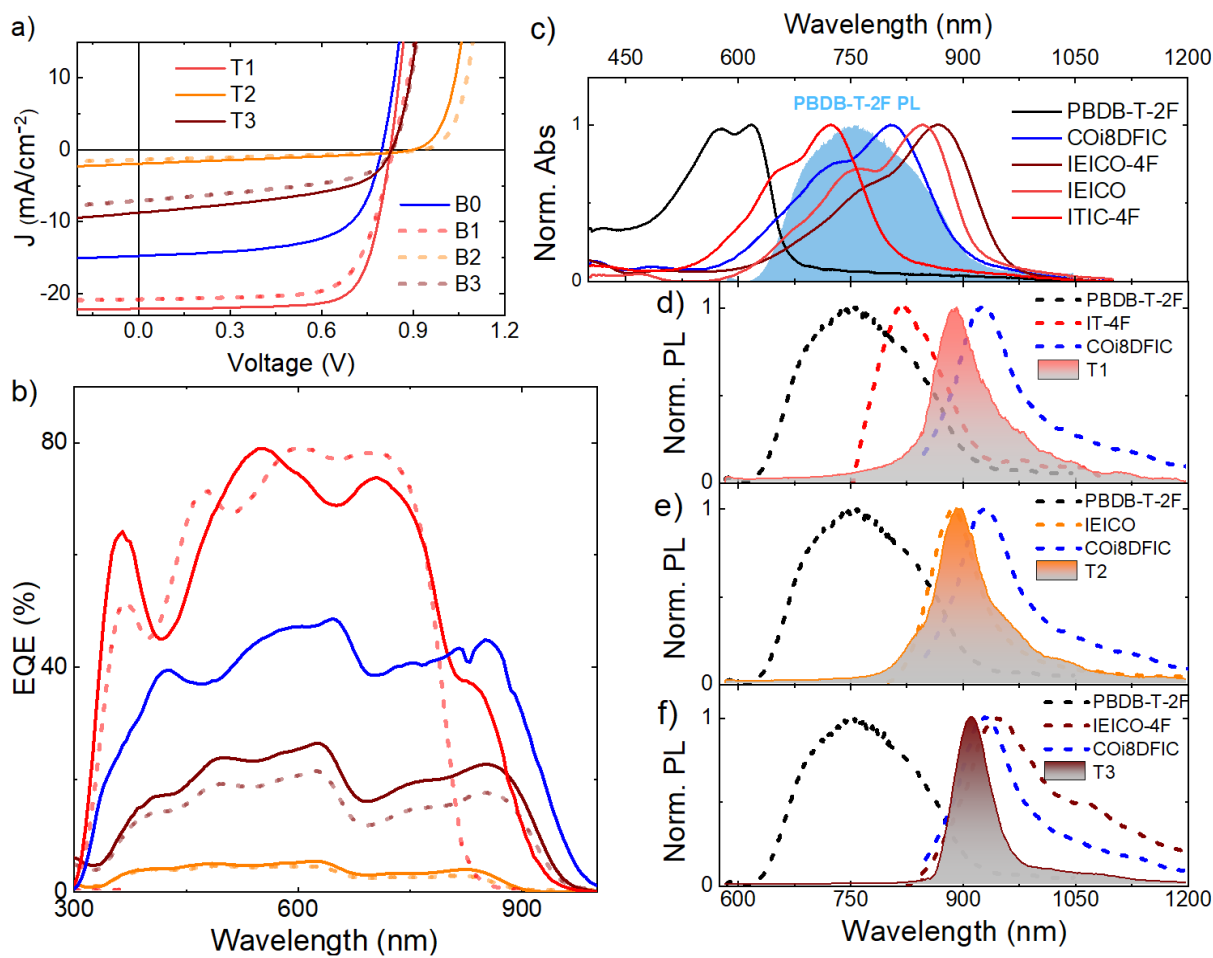

*Figure 2: **Photocurrent generation increases with the IE offset in the studied blends as donor excitation energy is transferred to the NFAs.** a) J-V curves of binaries: B0, PBDB-T-2F:COi8DFIC (1:1); B1, PBDB-T-2F:IT-4F (1:1); B2, PBDB-T-2F:IEICO (1:1); B3, PBDB-T-2F:IEICO-4F (1:1) and ternaries **T1**, PBDB-T-2F:IT-4F:COi8DFIC (1:0.9:0.1); **T2**, PBDB-T-2F:IEICO:COi8DFIC (1:0.9:0.1); and **T3**, PBDB-T-2F:IEICO-4F:COi8DFIC (1:0.9:0.1). b) External quantum efficiency (EQE) spectra of the corresponding devices. c) Normalized absorption spectra (continuous lines) of the neat films of the materials used and photoluminescence emission spectrum (shaded area) of PBDB-T-2F. d) Normalized photoluminescence emission of the ternaries and corresponding neat films. All the ternary films were excited at 500 nm. The neat films of PBDB-T-2F, IT-4F, COi8DFIC, IEICO, IEICO-4F were excited at 500 nm, 650 nm, 750 nm, 750 nm, and 800 nm, respectively.*

We fabricated binary and ternary OSCs using an inverted device architecture, precisely ITO/ZnO (sol-gel)/bulk heterojunction (BHJ)/MoO$_x$/Ag.[21] The photocurrent in the binaries increases with $\Delta IE$, in line with our previous study,[13] from PBDB-T-2F:IEICO (**B2**), PBDB-T-2F:IEICO-4F (**B3**), PBDB-T-2F:COi8DFIC (**B0**) to finally PBDB-T-2F:IT-4F (**B1**), see Figure 2a and Table S2. This is also reflected by the external quantum efficiency (EQE) spectra (Figure 2b), see also Figure S11 for the IQE spectra.

We prepared ternary BHJs by adding a small quantity of COi8DFIC acceptor, that is, BHJs composed of D:NFA-1:COi8DFIC with a weight ratio of 1:0.9:0.1. This corresponds to adding a lower *IE* NFA to IT-4F (**B1** to **T1**), a higher *IE* NFA to IEICO (**B2** to **T2**), and a slightly higher *IE* NFA to IEICO-4F, from **B3** to **T3**, respectively. The photocurrent in **T1**, **T2**, and **T3**, follows the same trend as in the binaries (**B1** to **B3**).

Comparing the ternaries to the corresponding binaries, the addition of the third component COi8DFIC increases the photocurrent in all three cases, but for different reasons. As shown by the EQE spectra (Figure 2b), in the large offset blend (**T1**, IT-4F), the photocurrent increase originates from the additional near-infrared absorption of COi8DFIC in the 800-900 nm spectral region (Figure 2b). In contrast, in the case of the intermediate offset (**T3**, IEICO-4F), the EQE clearly increases across the entire spectrum. Finally, only a minor improvement was observed in the low offset case (**T2**, IEICO). Below we examine the reasons for the differences in photocurrent generation improvements.

An improved morphology and packing have often been reported to enhance the charge transport[22] and reduce the recombination[4a, 22-23] in TSC. We thus conducted grazing incidence wide-angle X-ray scattering studies (GIWAXS) to track the evolution of the packing upon the addition of the third component.[24] The GIWAXS patterns obtained for the binaries and the ternaries display solely diffraction features from PBDB-T-2F aggregates (of similar intensity), indicating that the NFAs are mostly amorphous in all blends (Figure S15 of the Supplementary Information). These observations are in line with the previous studies.[25] Hence, we concluded that the observed differences in charge generation do not result from differences in the solid-state microstructure.

Turning now to the photophysics, we see in Figure 2c that the photoluminescence of PBDB-T-2F strongly overlaps with the absorption of the acceptors, thus being favorable to FRET.

Steady-state photoluminescence (ssPL) spectroscopy confirms the expected energy transfer as PBDB-T-2F photoexcitation leads to the emission from the NFAs (Figure 2d-f) in all three ternaries. We note that FRET is expected to have only an indirect effect on efficiency: FRET does not increase or decrease the efficiency, but rather makes it *IE*-dependent instead of being *IE*- and *EA*-dependent. Indeed, due to FRET, all the excitation ends up in the NFAs prior to charge transfer, independently of the material in which the light was initially absorbed. As a result, charge transfer is always a hole transfer from the acceptor to the donor, and as such unlikely to be influenced by the *EA* of the materials. This explains why IEICO and IEICO-4F-based blends offer poor and moderate quantum efficiencies, respectively, despite *EA* alignments very favorable to electron transfer from the donor to the NFAs.

Those preliminary measurements did not reveal any significant difference between the three ternary blends; we thus turned to transient absorption spectroscopy to investigate the differences in the charge generation and recombination processes.

## 2. Transient Absorption Spectroscopy

Ultrafast transient absorption spectroscopy (TAS) can reveal energy and charge transfer dynamics in TSC's photoactive layers. TAS allows identifying which of the blend's component bears the photoexcited states by monitoring the bleach of the materials absorption (ground state bleach, GSB) caused by the depopulation of the ground state of the molecules. Moreover, the different excited states (charges, excitons) can be distinguished by their characteristics photoinduced absorption (PA) features.[26] Finally, excited state dynamics can be monitored across the entire sub-picosecond to microsecond timescale following pulsed photoexcitation.

Following PBDB-T-2F excitation, ultrafast energy transfer is observed in the first few picoseconds in all three blends. More precisely, in **T1**, selective PBDB-T-2F photoexcitation (~505 nm) causes distinct COi8DFIC and IT-4F exciton-induced absorption peaking at 1.2 and 1.3 eV, respectively, as early as 400 fs after photoexcitation, as shown in Figure 3a. Similarly, **T2** and **T3** exhibit NFA exciton-induced signatures such as the NFAs ground state photo-bleach at 1.5 and 1.42 eV, respectively, observed as early as 300 fs after PBDB-T-2F photoexcitation, see Figures 3b and 3c. We note that, in **T2** and **T3**, the exciton-induced absorption band of IEICO (**T2**) and IEICO-4F (**T3**) overlap with that of PBDB-T-2F. We distinguish these features by probing the GSB bands as they are clearly separated (Figures 2b and 2c). On the other hand, the GSB of the NFAs in **T1** is superimposed by the PBDB-T-2F cation absorption (see Figure 3a), so it cannot be directly observed. A detailed assignment of all spectral features and their kinetics can be found in the Supporting Information (Figure S17).

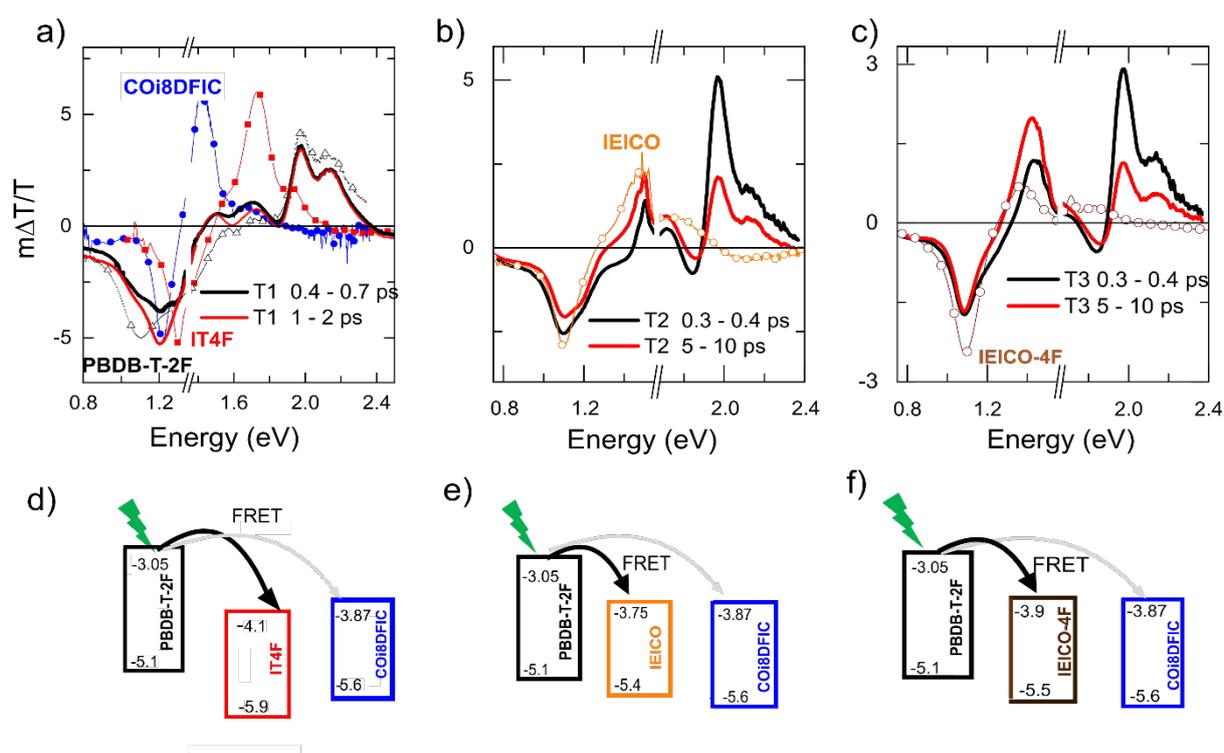

*Figure 3: **Ultrafast energy transfer from PBDB-T-2F revealed by the instantaneous occurrence of NFAs singlet excitons upon selective PBDB-T-2F excitation**. TA spectra of the first picoseconds following the selective excitation of PBDB-T-2F at 505 nm (solid line): (a) **T1**, (b) **T2** and (c) **T3**. Reference exciton spectra (line + symbol) of neat films are shown for comparison and were obtained by TA measurement probed 1 ps after excitation at 505 nm for PBDB-T-2F, 700 nm for IT-4F, and 850 nm for COi8DFIC . The x-axis break is due to the saturation of the detectors caused by the scattered fraction of the high intensity 800 nm pulse used to generate the probe white light. The fluence used was ~4 µJ/cm$^2$. (d-f) Schematic representation of energy transfer from PBDB-T-2F to acceptors.*

While energy transfer occurs similarly in the three blends, charge transfer is very system dependent: efficient and fast in **T1**, contrasting with a slower and less efficient charge generation in **T2** and **T3**. In **T1**, PBDB-T-2F cations are already detected in the first hundreds of femtoseconds, as indicated by charge-induced absorption from 1.4 – 1.8 eV (+ symbols in Figure 4a). This charge-induced absorption is seen indirectly by its effect of masking the GSB of IT-4F and COi8DFIC (1.4 – 1.8 eV), which should otherwise have a similar amplitude to that of the PA signal of the NFAs excitons and anions between 1.1 – 1.4 eV. In contrast, in **T2** and **T3**, no PBDB-T-2F charge signal is seen at early times (Figure 3b-c), here, the PA and GSB signatures of NFA excitons are observed, similar to the spectra of neat samples, i.e., in the absence of charges and presence of excitons.

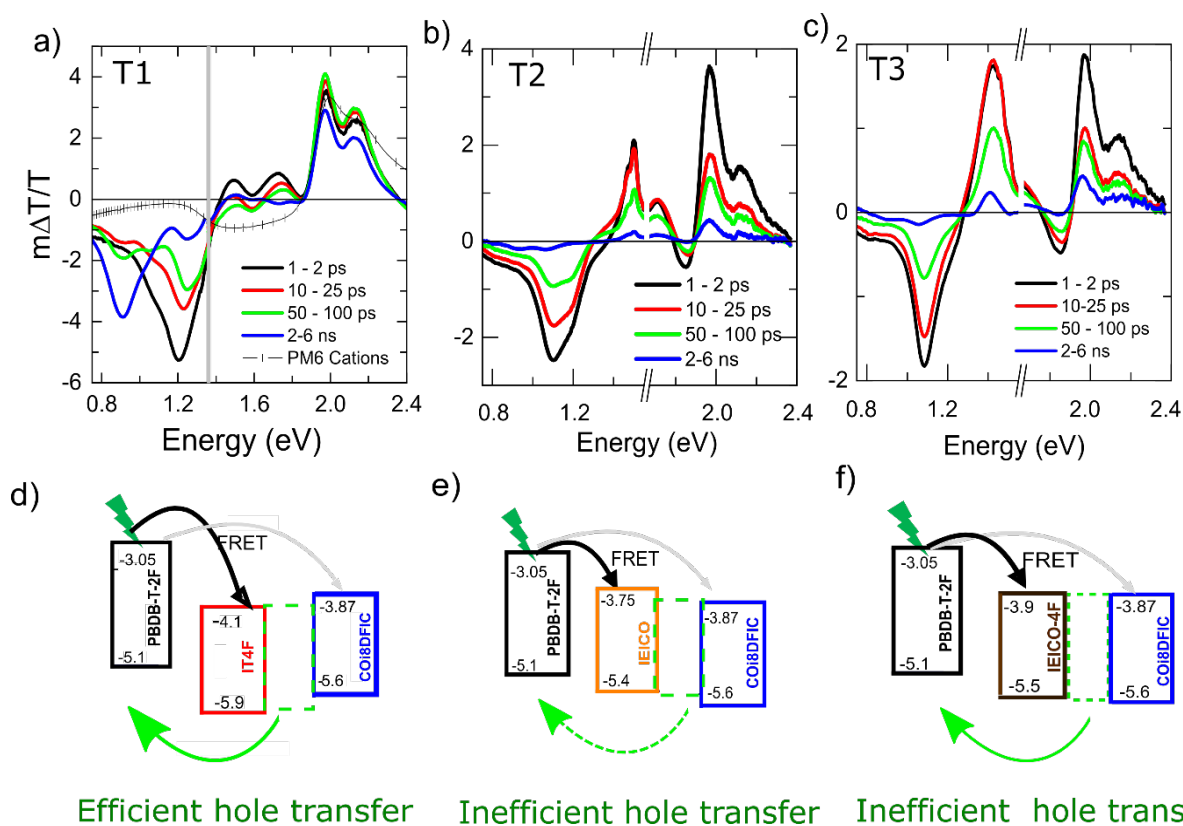

*Figure 4: **Evidence of efficient ultrafast charge generation in T1 contrasting with a slower and inefficient generation in T2 and T3.** Evolution of the ps-ns TA spectra following selective excitation of PBDB-T-2F for (a) **T1** ($\lambda_{pump}$= 505 nm), (b) **T2** ($\lambda_{pump}$= 550 nm), and (c) **T3** ($\lambda_{pump}$= 550 nm). The + symbol plot in a) represents the PBDB-T-2F cation spectra obtained by oxidation with FeCl$_3$. The vertical bar in (a) separates the measurements probed using two different detectors to cover a broader range. The horizontal breaks in (b) and (c) indicate the region*

The difference in the charge generation efficiency remains or even increases over the time range monitored (Figure 4). In **T1**, the PBDB-T-2F GSB seen from 1.9 – 2.3 eV remains significant in the nanosecond time range, indicating an efficient formation of long-lived states. On the contrary, the GSB of PBDB-T-2F decayed fast in **T2** and **T3**, and only a limited excited state signal remains after 500 ps. In all three cases, the spectral signatures of the long-lived states (seen by the comparison of the 50 - 100 ps spectrum of **T1** with the 500 – 800 ps spectrum of **T2** and **T3**) differ from those of the excitons. We attribute those to PBDB-T-2F cations and NFA anions generation. More precisely, the GSB of PBDB-T-2F increased from 1 ps to 30 ps in **T1** (Figure 4a and S14) due to diffusion-mediated hole transfer from the acceptor. We note that in **T1**, a new PA peak ~0.9 eV appears on the nanosecond time scale, which we attribute to COi8DFIC triplet formation via charge carrier recombination since it corresponds to the triplet induced absorption of COi8DFIC seen in the binary systems, Figure S17 and Figure S18 in the SI. It is also consistent with the increased weight taken by the COi8FDIC photobleach (1.45 eV) relatively to that of IT-4F (1.65 eV). We note that the tripet band is well pronounced, despite the charges remaining abundant in the blend which we tentatively attribute to the fact that triplets often have a much larger absorption cross-section than charge carriers.[26]

To summarize the TAS observations, FRET from PBDB-T-2F occurs equally well in the three blends and is directed toward both NFAs. On the other hand, charge transfer is very system-dependent and efficient in **T1** but much less for **T2** and **T3**, and we identify it as the limiting step for device performance.

Importantly, this charge transfer appears to work equally well (or equally bad) from both NFAs involved in the ternary: the charge generation is efficient in (**T1**) for both IT-4F and COi8DFIC, whereas in **T2** and **T3**, it is less efficient for both NFA-1 and COi8DFIC (see TAS under specific acceptors' excitation in Figure 5 and Figure S19). In other words, at this stage, it seems that the majority acceptor NFA-1 controls the charge generation efficiency of the minority acceptor COi8DFIC, that is, at the PBDB-T-2F:COi8DFIC interface, which is present in all three blends, as summarized in Figure 4d-f. As we will see later, the reality is a bit more complex.

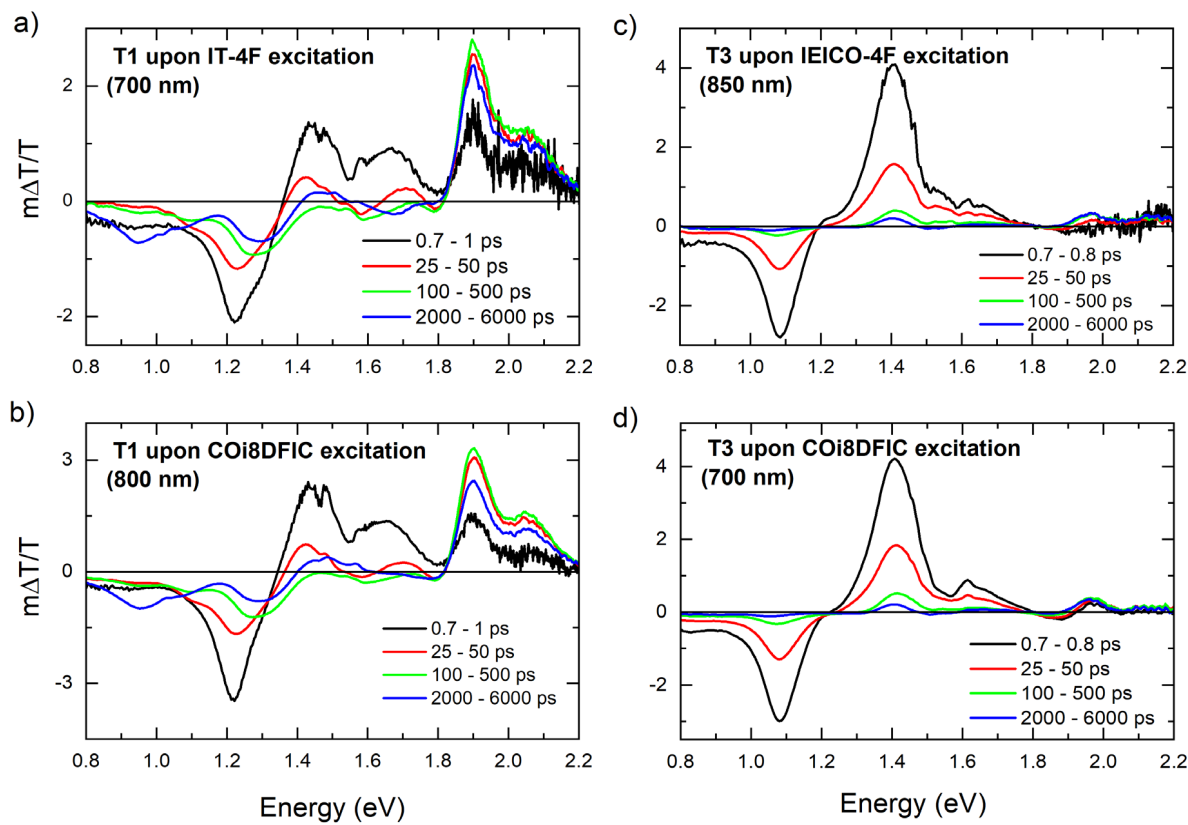

*Figure 5: The two acceptors work has a single channel for charge generation, and specific excitation of COi8DFIC can generate charges efficiently (b) or inefficiently (d) depending on the NFA-1 it is associated with: ps-ns TA spectra following selective excitation of (a) IT-4F in **T1** ($\lambda_{pump}$= 700 nm), (b) COi8DFIC in **T1** ($\lambda_{pump}$= 800 nm), (c) ) IEICO-4F in **T3** ($\lambda_{pump}$= 850 nm), COi8DFIC in **T1** ($\lambda_{pump}$ = 800 nm). For T2, the two acceptors have too close bandgaps, it is thus not possible to excite selectively one of the acceptors (700 nm and 850 nm excitations shown in Figure S19).*

Our observation is consistent with a *ΔIE*-driven hole transfer, where *ΔIE* is needed to overcome the energy levels bending at the D/A interface, as illustrated in Figure 6.[15] This energy level bending is mainly caused by the interaction of the charges with the surrounding molecules' quadrupole moments.[11, 13, 15, 27] A slight deviation from this trend, however: we would expect more difference in hole transfer efficiency between **T2** and **T3** due to the different *ΔIE* and the different photocurrents. A possible explanation is that PBDB-T-2F:IEICO-4F:COi8DFIC (**T3**) may have enough offset to generate some charges in short circuit conditions (in which the EQE is measured) due to the internal field, but not in open circuit conditions (in which the TAS is measured). This would make sense given that the *ΔIE* between PBDB-T-2F and the majority acceptor in that blend is very close to the reported average energy level bending <*B*> for PBDB-T-2F-based binaries.[13] This means that $IE^A_{interface}$ is generally very close to $IE^D_{interface}$ making the applied field contribution decisive in the relative position of those two levels, and hence the charge transfer.[13] The rather low fill-factor (49%, Table S2) of **T3**-based cells also supports the idea of a field-dependent generation, as it shows that moving away from the almost flat band conditions (at $V_{OC}$) to a larger internal field (towards short circuit condition) is needed to maximize the current. We note that TAS does show more long-lived species in **T3** than in **T2**, but not in the same proportions as the differences observed in EQE and $J_{SC}$. We further note that rates extracted from TAS have been largely proven to reproduce the devices performances in simulations.[7b, 26, 28] However, in the case of a strongly field-dependent charge generation, this dependence has to be measured separately, for example using time delayed collection field experiments, and added to the model, which is out of the scope of this work.[29]

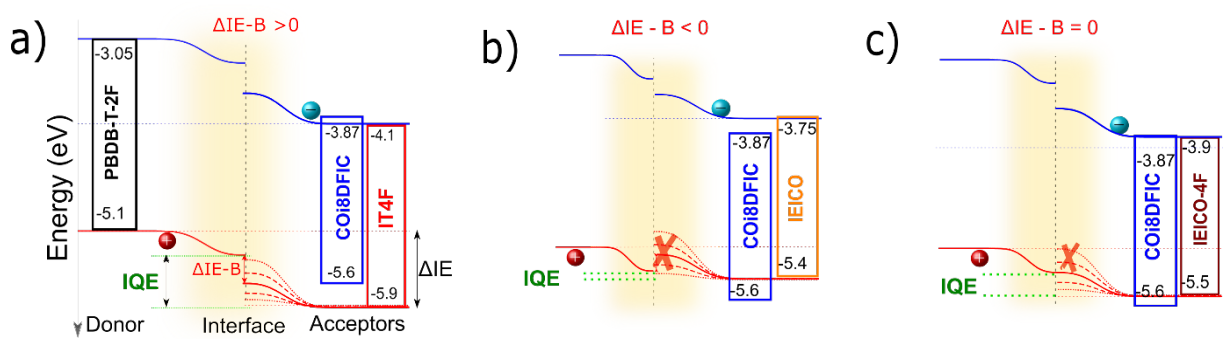

*Figure 6: **The interplay of the energy level bending and ionization energy offset at the donor-acceptor heterojunction controls the charge transfer efficiency.** Considering a Gaussian distribution of energy level bendings, the charge transfer efficiency corresponds to the fraction of the interfaces where the local energy levels bending is lower than ΔIE, i.e. (a) most of the D/A interfaces in **T1**, (b) a small fraction in **T2**, and (c) close to the*

*half of them in **T3** where the offset between the IE of PBDB-T-2F and the most abundant acceptor IEICO-4F is close to average energy level bending of 0.4 eV reported for PBDB-T-2F:NFA sysems.*[13]

## 3. The role of the third component in the charge generation

The TAS studies suggest that the IE of the majority NFA dictates the EQE of the TSC. However, the EQE spectra (see Figure 2b) do show that, in some cases, the third component has some effect other than just the additional absorption: the EQE of PBDB-T-2F:IEICO-4F:COi8DFIC (**T3**) is increased compared to PBDB-T-2F:IEICO-4F across the entire spectral region, rather than only in the COi8DFIC absorption region. On the other hand, this effect appears to be minimal in the other two blends, which showed only a minor improvement (**T2**), or no change at all (**T1**). To clarify the role of the minority NFA, NFA-2, we enhanced its impact by increasing the weight fraction of COi8DFIC in the ternary blends.

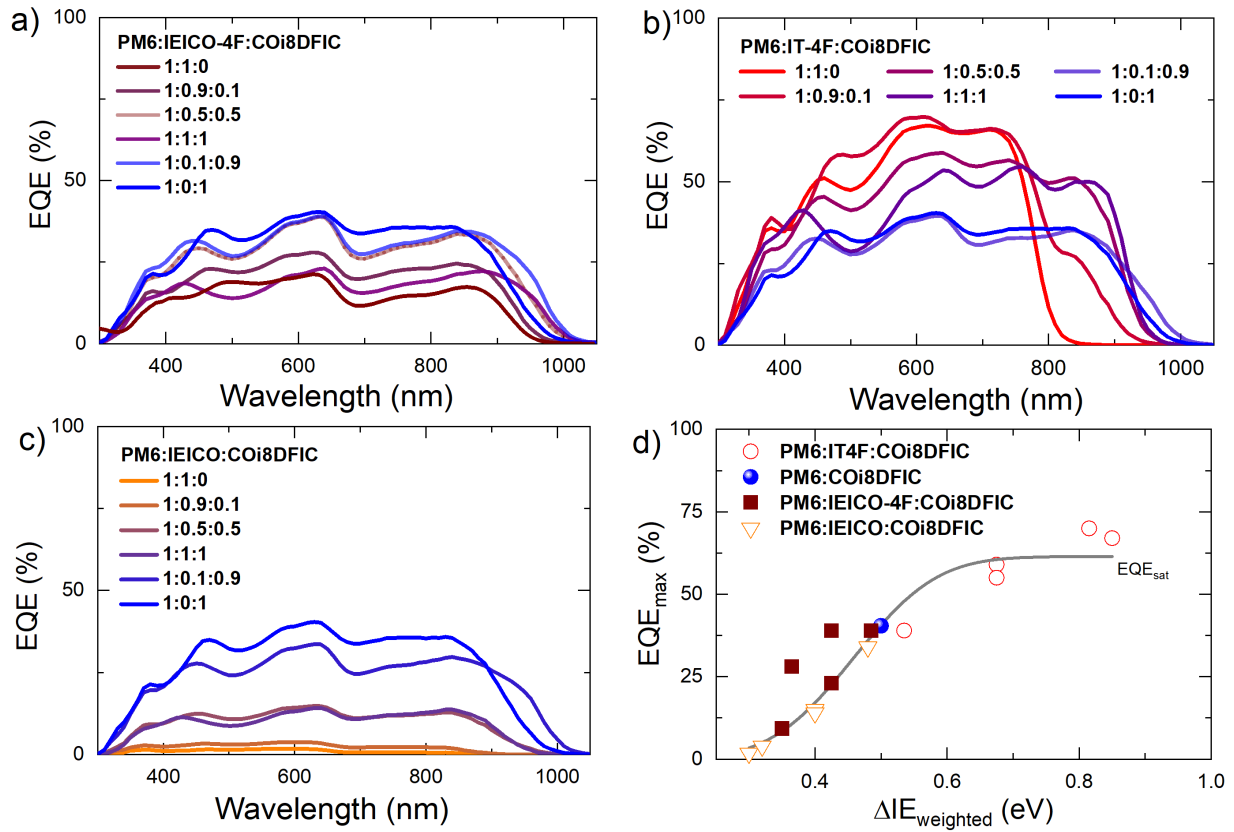

*Figure 7: **The quantum efficiency of ternary solar cells increases with the offset between the IE of the donor and the weighted average of acceptors IEs following the same law as the EQE of binary cells with the donor:acceptor IE offset**. EQE spectra of ternary and binary solar cells: a) PBDB-T-2F:IEICO-4F:COi8DFIC, b) PBDB-T-2F:IT-4F:COi8DFIC, and c) PBDB-T-*

*2F:IEICO:COi8DFIC for different NFAs ratios. The J-V curves and statistical data of PV parameters can be found in Table S2 and Figure S3- S10. d) EQE maximum vs weighted average IE offset: [wt%(NFA-1) × IE(NFA-1) + wt%(COi8DFIC) × IE(COi8DFIC)]. The QE can be well fitted (solid line) by the fraction of the interface which exhibits energy level bending lower than ΔIE, assuming a Gaussian distribution of energy level bendings with an average <B> and a standard deviation σ, we obtain $EQE = \frac{EQE_{sat}}{2}(1 + erf\frac{\Delta IE - <B>}{\sigma})$.[13] This fit yields <B> = 0.46 ± 0.02 eV, σ = 0.140 ± 0.03 eV and $EQE_{sat}$ = 61 ± 6 %.*

As shown in Figure 7 a-c, the EQE spectra evolves continuously with the NFA-1:COi8DFIC ratio in the ternaries from the PBDB-T-2F:NFA-1 binary to the PBDB-T-2F:COi8DFIC binary. Noticeably, the spectra appear relatively flat as long as the absorption is not limited by the too low weight fraction of one of the components. The spectra 'wave-shape' is most likely related to optical interference features caused by the multilayer structure of the solar cell. This flatness is also consistent with our observation of a single channel for charge generation.

This EQE spectrum flatness has two important consequences. First, we can safely use its maximum value $EQE_{max}$ as representative of the efficiency of internal processes: as the spectral variations are primarily due to the absorption spectrum of the active layer in the device structure, using $EQE_{max}$ enables to minimize the influence of absorption losses and focus on other photophysics. Second, the influence of energetics on solar cell efficiency can be represented by a single variable accounting for the energy levels of the three blends' components. In the frame of our previous findings for charge generation in binary blends,[12a] we chose $\Delta IE_{weighted}$: the difference between the ionization energy of the donor, and the average of the acceptors ionization energies is weighted by considering their weight ratio in the ternary blend. The result is shown in Figure 7d.

Figure 7d clearly shows that $EQE_{max}$ and $\Delta IE_{weighted}$ are correlated. We fitted the experimental data ($EQE_{max}$) with an equation derived from that observed for binary blends,[13] and obtained an excellent agreement (solid line):

$$EQE_{max} = \frac{EQE_{sat}}{2}(1 + erf\frac{\Delta IE_{weighted} - <B>}{\sigma}) \qquad (1)$$

As a reminder, this relation accounts for a Gaussian distribution of energy level bendings at the D/A interface, with an average <B> and a variance σ. The internal quantum efficiency is equal to the integration of this distribution from a bending of 0 eV to a bending equal to $\Delta IE$, multiplied by a prefactor $EQE_{sat}$. This function corresponds to the fraction of the energy level bending distribution for which the hole in the highest occupied molecular orbital (HOMO) of the donor at the interface is higher in energy than in the HOMO of the acceptor at the interface, making the hole transfer favorable (see Figure 1).

For the ternary blends represented here, the fit yielded an average bending $B$ = 458 ± 19 meV with a variance of σ = 141 ± 22 meV, slightly larger that has been reported for the binaries,[13] and an upper $EQE_{sat}$ limit of 61 ± 6 %. We note that the fit applied here describes the evolution of the quantum efficiency of the hole transfer process. Other losses, such as incomplete photon absorption and exciton decay to the ground state - note that the films exhibit some remaining PL (see Figure 2e) – are not accounted for, thus explaining that the $EQE_{sat}$ is limited to about 58%. The internal quantum efficiency spectra (IQE) were also calculated for the binary blends as well as the ternaries with blend ratios 1:0.9:0.1 (see Figure S11 – S12). Fitting eq. (1) to the spectrally averaged IQEs yields similar results (<B> = 454 ± 25 meV, σ = 140 ± 27 meV), but with an $IQE_{sat}$ of 86 ± 3%, confirming a large part of the missing EQE is due to absorption losses (see Figure S13).

From an energetic perspective, the mixture of NFA-1 and COi8DFIC thus appears to behave as a single material whose *IE* can be described by the linear combination of their *IE*s. Such behavior has previously been predicted theoretically and observed experimentally on co-evaporated acceptor molecules.[11-12] It quantitatively explains the evolution of hole transfer recently reported for ternary solar cells based on one donor and two acceptors.[30] Note that the *IE* of the NFA mixture corresponds to the weighted average of both materials *IE*s, rather than to the *IE* of a brand new state with unique electrostatic interactions. As such, we do not consider that the mixture qualifies as an alloy.[30-31]

This description readily explains why a difference was observed between the QE of the binary blend PBDB-T-2F:IEICO-4F and that of the ternary PBDB-T-2F:IEICO-4F:COi8DFIC but not in the two other cases (see Figure 2). Indeed, the *ΔIE* between PBDB-T-2F and IEICO-4F as well as between PBDB-T-2F and COi8DFIC are very close to <*B*> the maximum of the Gaussian distribution of energy level bendings. As such, even the slight increase of *ΔIE*$_{weighted}$ caused by adding 10% of COi8DFIC to the blend makes a significant fraction of the energy level bendings smaller than *ΔIE*$_{weighted}$, thus making those interfaces able to generate charges. On the other hand, the *ΔIE* of PBDB-T-2F:IEICO and PBDB-T-2F:IT-4F are on the lower and higher plateaus of the error function, respectively, and are thus relatively unaffected by small changes in *ΔIE*$_{weighted}$.

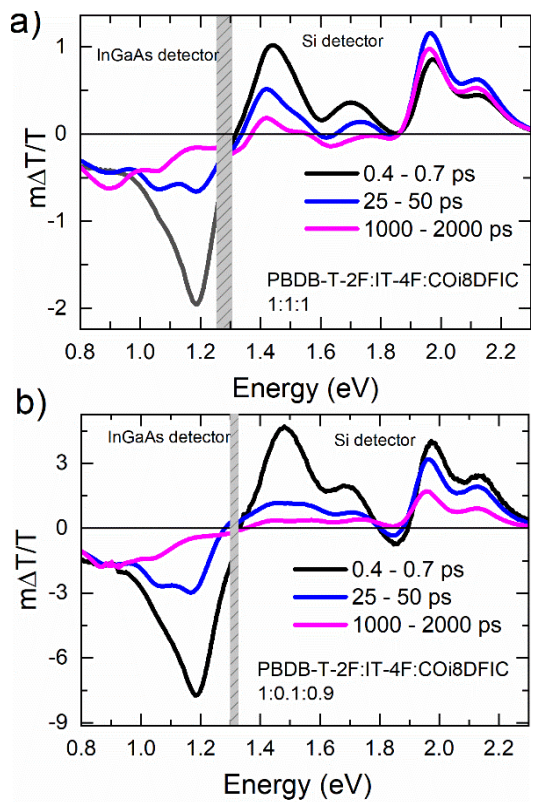

*Figure 8: **The quantity of charges generated in PBDB-T-2F:IT-4F:COi8DFIC decreases upon increasing the fraction of COi8DFIC.** Evolution of the TA spectra in the ps-ns timescale following the selective excitation of PBDB-T-2F at 505 nm for PBDB-T-2F:IT-4F:COi8DFIC with ratios (a) 1:1:1 (~4μJ/cm$^2$) and (b) 1:0.1:0.9. (~8 μJ/cm$^2$). The grey vertical bar between 1.35-1.4 eV separates the TA spectra measured with two different detectors, as indicated in the figure.*

TAS confirms that the evolution of the blends' EQE with the blend ratio originates from the hole transfer efficiency. TAS measurements on PBDB-T-2F:IT-4F:COi8DFIC ternary films (Figure 8) reveal fast FRET for the 1:1:1 and 1:0.1:0.9 blends, evidenced by instantaneous NFA exciton-induced absorption signals upon selective PBDB-T-2F excitation. However, the intense PBDB-T-2F cation absorption signal (1.4 – 1.8 eV) that was previously seen in the 1:0.9:0.1 blend at early delay times (Figure 3a and Figure 4a) is not observed here. Moreover, in contrast to the 1:0.9:0.1 ternary blend, where almost all initial PBDB-T-2F photoexcitations forms long-lived states, in the 1:1:1 blend film, a small decrease of the PBDB-T-2F GSB (1.9 – 2.2 eV) signal was observed within the formation time of long-lived species. But in the 1:0.1:0.9 blend, more than 50% of the signal decays before forming long-lived species (sub-nanosecond timescale). This indicates that an equivalent fraction of the initial photo excitations was not converted into charges in these blends.

A practical consequence of charge transfer in ternary blends being controlled by $\mathit{\Delta IE}_{\text{weighted}}$, is that we can activate this charge transfer in D/A blends that would otherwise have a too low $\mathit{\Delta IE}$, by adding a third component with a deeper $\mathit{IE}$. We selected PBDB-T-2F:IEICO with an $\mathit{IE}$ offset ~0.3 eV and thus EQE below 5% (See Figure 2) and added a third component, IT-4F, with a deeper $\mathit{IE}$ which offers an offset of 0.8 eV with PBDB-T-2F. As shown in Figure 7a, the EQE in the spectral region of IEICO absorption increases from close-to-zero to more than 25% upon adding 40 wt.% IT-4F. Here we note that these ternary devices are not optimized but are used here as a proof of concept. Any further increase is impeded since, decreasing the blend's IEICO fraction, the limited absorption in this spectral region becomes a bottleneck.

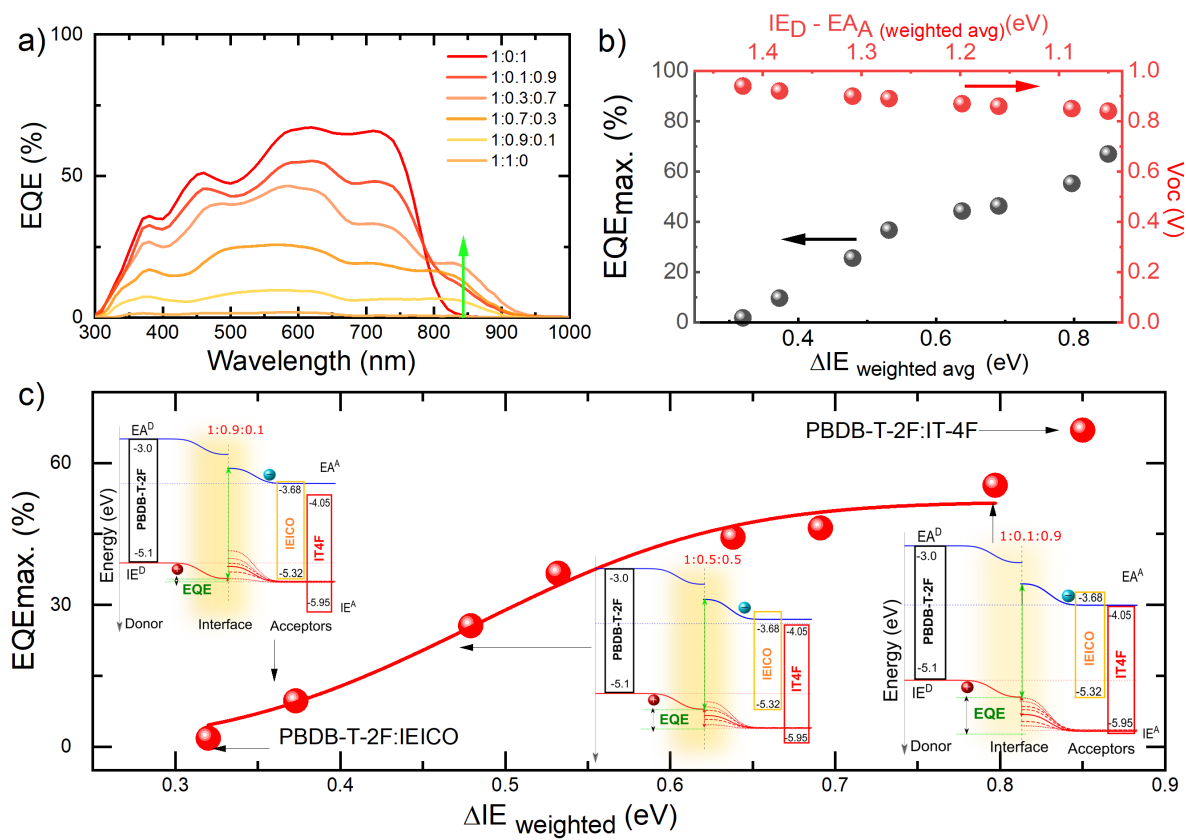

*Figure 9: **Charge transfer in a otherwise too low IE offset system (here PBDB-T-2F:IEICO) can be activated by adding a deep IE third component, here IT-4F**. a) EQE spectra of PBDB-T-2F:IEICO:IT-4F blends for different IEICO:IT-4F ratios. b) EQE maximum vs weighted average IE offset for PBDB-T-2F:IEICO:IT-4F solar cells (black) and the evolution of $V_{OC}$ with the difference between the donor's IE and the weighted average acceptor EA (red). c) The error function fit to the evolution of EQE maximum vs weighed average IE offset yield B = 0.39 eV and σ = 0.08 eV. Insets show the schematic of the energy level bending at the donor–acceptor heterojunction due to the electrostatic interaction of charges with the quadrupole moments of the surrounding molecules.*

We observe that $EQE_{max}$ for these IEICO:IT-4F ternaries obeys the same trend found previously for binary donor:NFA blends[13] (Figure 9c) as well as reported here for the other ternary systems. The insets in Figure 9c schematically represent the proposed interfacial energetics for a selection of different blending ratios. The corresponding device J-V curves can be found in SI (Figure S20).

Finally, we tested the generality of this approach on two other blends using very different molecules: one based on Y6 (PBDB-T-2F:Y6:IEICO) and one using only non-fluorinated acceptors (PBDB-T-2F:IEICO:IDIC). As shown in Figures S19 to S21, the same evolution is again obtained. However, PBDB-T-2F:IEICO:IDIC shows a QE efficiency increase only at larger *ΔIE* values (the fit finds *<B>* = 0.6 eV), which could be an interesting starting point for further studies.

Those results demonstrate that the mixing of two acceptors' ionization energies can be used effectively to continuously adjust the energy levels alignment between the donor(s) and the acceptor(s) to satisfy the previously identified criteria of *ΔIE* > *<B>* + 2σ that maximizes charge transfer.[13] This is a very interesting alternative to the use of molecular engineering by modifying existing molecules or synthesizing new ones. We stress that using this approach, *IE* of the pair of NFA can be adjusted continuously between the two NFAs *IE*s, which is not possible through molecular engineering. Finally, these findings pave the way to the easier use of low bandgap acceptors, which would overwise suffer from either low *ΔIE* with the donors impeding efficient charge generation or low difference between their *EA* and the *IE* of the donor, leaving only little room for the open circuit voltage of resulting cells.

As an opening to future studies, we note that the $V_{OC}$ also linearly evolves between the two extreme ratios (Figure 9b), suggesting that the acceptors *EA*s may also be mixed similarly to their *IE*s. This linear evolution of $V_{OC}$ is in agreement with previous reports.[10] The $V_{OC}$ change is, however, only 40 meV upon changing the average *EA* by more than 0.3 eV, stressing that further in-depth studies are needed to learn precisely about those energy levels mixing and about energy losses in organic solar cells in general.[32]

## Conclusion

We have investigated the energetics driving the quantum efficiency of ternary solar cells composed of one donor and two NFAs. The J-Vs showed a clear improvement of the current upon adding the third component in all three cases; the EQE spectra further showed that in PBDB-T-2F:IT-4F:COi8DFIC this improvement was primarily due to the additional absorption feature, whereas for PBDB-T-2F:IEICO:COi8DFIC, and PBDB-T-2F:IEICO-4F:COi8DFIC a QE improvement was observed over the whole spectral range. Photoluminescence and transient absorption measurement revealed efficient energy transfer from the donor to the acceptors in all three cases, implying that charge transfer mostly occurs from the acceptor, hence being a hole transfer controlled by their ionization energies.

Additionally, transient absorption showed that in the 1:0.9:0.1 blends, the two D:NFA interfaces behave identically as a single channel through which the hole transfer efficiency seems at first to be controlled by the offset between the donor and the acceptor present in the largest weighted fraction. However, varying the blend ratios revealed that the EQE and the charge transfer efficiency follow the offset between the *IE* of the donor and the average *IE* of the acceptors, weighted for their blending ratio. We found that the evolution of the EQE with the D/A *IE* offset can be described by the same equation as previously reported for binary solar cells, provided that the acceptor's *IE* is replaced by the weighted average of the acceptors *IE*s.

This has important implications: it enables the mixing of two acceptors to continuously adjust the energetics of a D/A heterojunction to maximize the quantum efficiency while minimizing the energy losses. This also paves the way for the use of NFAs that would otherwise suffer from their too limited *IE* offsets with the donor (weak photocurrent generation) or too deep *EA* (low $V_{OC}$). This is especially relevant for low bandgap materials for which the proximity of the IE and *EA* would otherwise offer minimal freedom for adjustment.

# Experimental Procedures

## Resource availability

### Lead Contact

Further information and requests for resources should be directed to and will be fulfilled by the lead contact, Julien Gorenflot (julien.gorenflot@kaust.edu.sa).

### Materials Availability

This study did not generate new unique reagents.

### Data and Code Availability

The data presented in this article will be available at figshare.com (corresponding DOI to be added).

The Matlab code used for the TAS data evaluation is available publicly on Github ( https://github.com/TheOldGogo/Matlab_Scripts ). The code was used to correct the chirp, concatenate data from several experiments, subtract the background noise and/or laser scattering, extract selected time delay or spectral slices from the data matrix and plot them.

## Molecules

Full material names: IT-4F, 3, 9-bis(2-methylene-((3-(1, 1-dicyanomethylene)-6, 7-difluoro)-indanone))-5, 5,11, 11-tetrakis(4-hexylphenyl)-dithieno[2, 3-d:2', 3'-d']-s-indaceno[1, 2-b:5, 6-b']dithiophene; Y6, (2, 20-((2Z,20Z)-((12, 13-bis(2-ethylhexyl)-3, 9-diundecyl-12, 13-dihydro-[1, 2, 5] thiadiazolo[3, 4-e] thieno[2, "30":4', 50] thieno[20, 30:4, 5] pyrrolo[3, 2-g] thieno[20, 30:4, 5] thieno[3, 2-b]indole-2, 10-diyl)bis(methanylylidene))bis(5, 6-difluoro-3-oxo-2, 3-dihydro-1H-indene-2, 1-diylidene))dimalononitrile). IDIC: indacenodithiophene end capped with 1, 1-dicyanomethylene-3-indanone; IEICO-4F, 2, 2'-[[4, 4, 9, 9-tetrakis(4-hexylphenyl)-4, 9-dihydro-s-indaceno[1, 2-b:5, 6-b']dithiophene-2, 7-diyl]bis[[4-[(2-ethylhexyl) oxy]-5, 2-thiophenediyl]methylidyne(5, 6-difluoro-3-oxo-1H-indene-2, 1(3H)-diylidene)]]

bis[propanedinitrile]; IEICO, 2, 2' ‐ ((2Z, 2'Z) ‐ ((5, 5' ‐ (4, 4, 9, 9 ‐ tetrakis(4 ‐ hexylphenyl) ‐ 4, 9 ‐ dihydro ‐ s ‐ indaceno[1, 2 ‐ b:5, 6 ‐ b']dithiophene ‐ 2, 7 ‐ diyl)bis(4 ‐ ((2 ‐ ethylhexyl) ‐ oxy)thiophene ‐ 5, 2 ‐ diyl))bis(methanylylidene))bis(3 ‐ oxo ‐ 2, 3 ‐ dihydro ‐ 1H ‐ indene ‐ 2, 1 ‐ diylidene))dimalononitrile; PBDB-T-2F, poly[(2, 6-(4, 8-bis(5-(2-ethylhexyl-3-fluoro)thiophen-2-yl)-benzo[1, 2-b:4, 5-b']dithiophene))-alt-(5, 5-(1', 3'-di-2-thienyl-5', 7'-bis(2-ethylhexyl)benzo[1', 2'-c:4', 5'-c']dithiophene-4, 8-dione)];

IT-4F, Y6 and IDIC, were purchased from Solarmer, IEICO-4F and COiDFIC were purchased from 1-material. PBDB-T-2F, IEICO and IEICO were purchased from 1-material except the batches used from the complementary experiments on PBDB-T-2F:Y6:IEICO and PBDB-T-2F:IDIC:IEICO which were purchased from Solarmer. The materials were used without further purification.

### OSC Device Fabrication and Measurements:

Pre-patterned Indium Tin Oxide (ITO) fingered glass substrates were cleaned in an ultrasonic bath with acetone and isopropyl alcohol for 8 min. each and treated with oxygen plasma for 10 min. A ZnO sol-gel interlayer was deposited by spin coating at 4000 rpm for 40 s, followed by annealing for 10 min at 200 °C in air. The recipe can be found elsewhere.[21] The PBDB-T-2F, IT-4F, and IEICO, COi8DFIC, IEICO-4F and IEICO were dissolved (20 mg/ml) in chlorobenzene overnight at 80 ºC. The appropriate ratios of each were stirred for a couple of hours before spin-coating the active layer. Active layers were spin-coated at 2000 rpm inside a glove box. The thickness of the active layers is approximately 100 nm. Finally, 10 nm $MoO_x$ followed by 100 nm silver were thermally evaporated at $10^{-6}$ bar. J–V curves were measured using a Keithley 2400 source meter and a WaveLabs sinus-70 solar simulator calibrated to 1 sun, AM1.5 G inside the glove box. The curves were measured from negative to positive bias with a dwell time of 10 ms and step size of 20 mV. The samples were measured in a glovebox that was maintained in a nitrogen environment at room temperature with $O_2 < 10$ ppm and humidity < 0.1 ppm. The EQE was measured using an integrated system from Enlitech, Taiwan. The measured $J_{sc}$ is within 5-10% of the integrated $J_{sc}$.

Internal Quantum Efficiency (IQE) spectra calculations:

The IQE of the devices is calculated as:

$$IQE = \frac{EQE}{1 - R - parasic.abs} \qquad (2)$$

The EQE spectra was measured on the best working pixel of the device with an integrated system from Enlitech, Taiwan. The Reflectance $R$ was measured on a fully ITO coated (un-patterned) glass substrate on which the same layers stack was deposited as on a normal device, the Ag layer was also evaporated on the whole device. A fully sputtered ITO was used to avoid the fringe effects of the ITO fingers during the reflectance spectra measurement using the integrating sphere of a Perkin Elmer 9500 absorption spectrometer. The parasitic absorption spectra were obtained by transfer matrix modelling of the device stack, using the MATLAB code developed by Burkhard, Hoke and McGehee and available at https://web.stanford.edu/group/mcgehee/transfermatrix/ .[33] The active layer refractive indexes were measured in-house by variable angle spectroscopic ellipsometry (VASE) using an M-2000 ellipsometer (J.A. Woolam Co., Inc) following the procedure described elsewhere,[34] while for the other layers reference refractive indexes were used. The reflectance and parasitic absorption spectra used to estimate the IQEs can be found in figure S12 and S24.

We note that due to the fact that the reflectance R comes from a different sample, and the parasitic absorption comes from simulation, they may not perfectly correspond to what happens in the samples where the sample where the EQE spectra are measured. We call this mismatch "optical correction error". As a results, our estimates of the IQE spectra can be written as:

$$IQE_{estimate}(\lambda) = IQE + optical\ correction\ error\ (\lambda) \qquad (3)$$

Our experience shows that the optical correction error tends to adopt an interference-like pattern, being alternatively positive (constructive) and negative (destructive interference).

In cases where the actual IQE is close to 100% the optical correction error can thus make $IQE_{estimate}$ ($\lambda$) exceed 100% at some wavelengths.

However, due to those successive negative and positive features, the optical correction error averages out upon spectral averaging. On the over hand, the actual IQE tends to be wavelength-independent due to the lack of thermal activation for photocurrent generation,[35] except in case of deeply subbandgap absorption,[36] and FRET that makes donor and acceptor excitation almost equivalent.[13]

We thus obtain the IQE value by spectral averaging of the $IQE_{estimate}$ ($\lambda$):

$$\begin{aligned}<IQE_{estimate}(\lambda)>_\lambda &=<IQE>_\lambda +<optical\ correction\ error\ (\lambda)>_\lambda \\ &= IQE + 0 = IQE\end{aligned} \qquad (4)$$

### UV-Vis-NIR absorption spectroscopy:

The UV-Vis-NIR absorption spectra were measured with a Cary 5000 UV-Vis-NIR spectrophotometer.

### Reference Charge Spectra:

The PBDB-T-2F film for cation spectrum measurement was obtained by oxidizing the PBDB-T-2F layer with $FeCl_3$ inside a glove box. Thin layer of PBDB-T-2F was spin-coated onto a UV treated quartz substrate with spin speed of 2500 rpm for 60 s, followed by annealing on a hot plate at 100 C° for 10 min. After cooling down to room temperature, the $FeCl_3$ was spin casted over it with the same speed. The active layers were obtained by dissolving the PBDB-T-2F in chlorobenzene and $FeCl_3$ in isopropyl alcohol, yielding solutions with concentrations of 10mg/ml and 5mg/ml, respectively.

The transmittance spectra of both neat ($T_n$) and oxidized ($T_{ox}$) films were measured in a nitrogen-filled custom-made sample chamber (with the help of a home-built sample holder) and the excited species spectra sTA calculated as the relative transmittance change:

$$sTA = \frac{T_{ox} - T_n}{T_n} \quad (5)$$

where, $T_{ox}$ and $T_n$ are the transmittance of the oxidized and the neat film, respectively.

### Steady-state PL spectroscopy:

The steady-state PL emission spectra were measured with a Jobin Yvon Fluorolog spectrofluorometer from Horiba. The respective films were spin-coated inside glove box on a cleaned quartz substrate. The cleaning procedure is mentioned in the OSC device and measurements section. The monochromatic beam was incident at an angle of 60° to the film and emission was collected at the same angle. The films were excited using Xe lamp source and detected using PMT-Si detector and/or InGaAs detector subjected to the emission range of NFA. To verify the energy transfer from PBDB-T-2F to acceptors, PBDB-T-2F was excited at 500 nm in all blend films (binary and ternary). The films were excited at different wavelengths as mentioned in the different parts of text. The emission spectra were corrected for the setups spectral sensitivity corresponding to the used configuration (detectors and gratings).

### Grazing-incident wide-angle X-ray scattering (GIWAXS):

The active layer films were spin coated on Si substrate under the same spin speeds as device. Si substrates were cleaned by ultra-sonicating in IPA and acetone consecutively for 8 minutes each. Further cleaned in oxygen-plasma for 5 minutes to remove any possible organic dust.

GIWAXS measurements were performed at the BL11 NCD-SWEET beamline at ALBA Synchrotron Radiation Facility (Spain). The incident X-ray beam energy was set to 12.4 eV using a channel cut Si (1 1 1) monochromator. The angle of incidence $\alpha_i$ was set between 0.1-0.15° to ensure surface sensitivity. The scattering patterns were recorded using a Rayonix® LX255-HS area detector, which consists of a pixel array of 1920 × 5760 pixels (H × V) with a pixel size of 44 × 44 μm$^2$.

Data are expressed as a function of the scattering vector ($q$), which was calibrated using $Cr_2O_3$ as standard sample, obtaining a sample to detector distance of 201.17 mm. The exposure times of the samples were 5s. 2D GIWAXS patterns were corrected as a function of the components of the scattering vector. Edges of the samples were removed to eliminate edge effects in the GIWAXS pattern.

Transient absorption spectroscopy (TA):

TA spectroscopy was carried out using a custom pump–probe setup. The output of a titanium:sapphire amplifier (Coherent LEGEND DUO, 4.5 mJ, 3 kHz, 100 fs) was split into three beams (2, 1, and 1.5 mJ). One of them was used to produce a white-light supercontinuum from 550 to 1700 nm by sending the 800 nm pulses through a sapphire (3 mm thick) crystal. The other two beams were used to separately pump two optical parametric amplifiers (OPA) (Light Conversion TOPAS Prime). TOPAS 1 generates tunable pump pulses, while TOPAS 2 generates signal (1300 nm) and idler (2000 nm) only. The signal from TOPAS 2 was used to produce a white-light super continuum from 350 to 1100 nm by being focused near the surface of a calcium fluoride ($CaF_2$) crystal which was mounted on continuously moving stage to slow down its degradation. The pump-probe delay was adjusted by reducing the pump beam pathway between 5.12 and 2.6 m while the probe pathway length to the sample was kept constant at ~5 m between the output of TOPAS 1 and the sample. The pump-probe path length was varied with a broadband retroreflector mounted on an automated mechanical delay stage (Newport linear stage IMS600CCHA controlled by a Newport XPS motion controller), thereby generating delays between pump and probe from −400 ps to 8 ns.

The samples (films on quartz substrates) were kept under vacuum ($10^{-6}$ mbar) during the entire measurements. For the spectra reported here, the excitation wavelength used was 550 nm for all ternary blend films (other excitation wavelengths shown in the supporting information) and the neat NFAs and PBDB-T-2F were excited near their absorption maxima. The transmitted fraction of the white light was guided to a custom-made prism spectrograph (Entwicklungsbüro Stresing) where it was dispersed by a prism onto a 512 pixels complementary metal-oxide semiconductor (CMOS) linear image sensor (Hamamatsu G11608- 512DA) for the *broadband* range or a 512 pixels negative channel metal oxide semiconductor (NMOS) linear image sensor (Hamamatsu S8381-512) for the *visible* range. The probe pulse repetition rate was 3 kHz, while the excitation pulses were directly generated at 1.5 kHz frequency, and the detector array was read out at 3 kHz. Adjacent diode readings corresponding to the transmission of the sample after excitation and in the absence of an excitation pulse were used to calculate $\Delta T/T$. Measurements were averaged over several thousand shots to obtain a good signal-to noise ratio. The chirp induced by the transmissive optics was corrected with a custom Matlab script. The delay at which pump and probe arrive simultaneously on the sample (i.e., zero time) was determined from the point of the maximum positive slope of the TA signal rise for each wavelength.

## Ultraviolet photoelectron spectroscopy (UPS) and low energy inverse photoelectron spectroscopy (LE-IPES):

*UPS* measurements on thin films of organic semiconductor were performed in ultrahigh vacuum chamber (base pressure of $10^{-10}$ mbar) equipped with a Sphera II EAC 125 7-channeltron electron analyzer, calibrated with the Fermi edge of clean polycrystalline silver. The measurements were performed using the He I line (excitation energy of 21.22 eV) at pass energy of 10 eV, with -10 eV of an external bias. The experimental uncertainty in the measurements is ± 0.05 eV.

*LE-IPES* measurements were performed in isochromatic mode[37] using an ultra-high vacuum (base pressure $10^{-9}$ mbar) set-up build in-house. The emitted photons were detected using a solid-state PMT detector (Hamamatsu R585) mounted outside of vacuum and equipped with a nm bandpass filter (Semrock) with a 10 nm wavelength window. Samples were measured immediately after the UPS measurements by transferring to the LE-IPES manipulator without air-exposure.

The ionization energy and the electron affinity of materials was determined from the UPS and LE-IPES spectra as described previously.[38] Samples for UPS/LE-IPES were prepared by spin coating a thin film of the organic semiconductor (polymer/NFAs) on a clean silicon substrate sequentially coated with a 10 nm Ti and 100 nm of sputtered gold. The onset energies of the occupied and unoccupied frontal molecular orbitals were derived by de-convoluting the spectra using Gaussian functions and a Taugaard background. The peaks fits are only used as a guide for precisely calculating the onset energies, and the attribution of deconvoluted peaks to specific orbitals need complimentary calculations.

## Supporting Information

Document S1. Supporting Information, containing Scheme 1, Figures S1-S16, and Tables S1-S2 in the following sections: Molecular Structures, Absorption spectra, UPS and LE-IPES spectra, Solar cell figures of merit and statistics for the PBDB-T-2F:NFA-1:COi8DFIC blends, Grazing incidence wide-angle X-ray scattering, PL Spectra, Additional transient absorption spectra and kinetics, Solar cell figures of merit for the PBDB-T-2F:IT-4F:IEICO blends.

## Acknowledgments

This publication is based on work supported by the KAUST Office of Sponsored Research (OSR) under award numbers OSR-2019-CARF/CCF-3079 and OSR-CRG2018-3746. J.M. and N.R.-G. thank the support from Xunta de Galicia ED431F 2021/09. GIWAXS experiments were performed at NCD-SWEET beamline at ALBA Synchrotron with the collaboration of ALBA staff.

## Author Contributions

S.K. performed the TAS experiment and wrote the first draft as well as large parts of the successive versions. S.H.K.P. processed, optimized and electrically characterized the devices for other blend ratios and wrote parts of the manuscript. A.S. measured the UPS and IPES. H.Y. optimized the original devices as well as the original films for TAS (1:0.9:0.1 ratios) and measured their absorption spectra under supervision of G.L. C.S.P.D.C performed and corrected some PL measurements, reviewed and edited the manuscript and participated in understanding the physics. S.C. calculated some of the IQE spectra. H.X. prepared and measured some of the complimentary solar cells. N.A. prepared samples and measure the spectra for the reference charge and triplet signals. N.R. performed the GIWAXS experiment and analysis under the supervision of J.M. J.K. performed some PL measurements. D.B.

supervised S.H.K.P. and A.S. F.L. supervised S.K., J.G., C.S.P.D.C, N.A., S.I. and J.K. J.G. led the data interpretation, coordinated the study, modeled the experimental behavior, and wrote parts of the manuscript. All authors contributed to the final draft.

## Declaration of interests

The authors declare no competing interests.

# Supporting Information

**Rationalizing the influence of tunable energy levels on quantum efficiency to design optimal non fullerene acceptor-based ternary organic solar cells.**

Safakath Karuthedath,[1,7] Sri H. K. Paleti,[1,7] Anirudh Sharma,[1] Hang Yin,[2] Catherine S. P. De Castro,[1] Si Chen,[1] Han Xu,[1] Nisreen Alshehri,[1,3] Nicolas Ramos,[4] Jafar I. Khan,[1] Jaime Martin,[4,5] Gang Li,[2] Frédéric Laquai,[1] Derya Baran[1,*] and Julien Gorenflot[1,6,**]

[1]King Abdullah University of Science and Technology (KAUST), KAUST Solar Center (KSC), Physical Sciences and Engineering Division (PSE), Material Science and Engineering Program (MSE), Thuwal 23955-6900, Kingdom of Saudi Arabia

[2]Department of Electronic and Information Engineering, The Hong Kong Polytechnic University, Hung Hom, Kowloon, Hong Kong SAR, P. R. China

[3]Physics and Astronomy Department, College of Sciences, King Saud University, Riyadh 12372, Kingdom of Saudi Arabia

[4]POLYMAT, University of the Basque Country UPV/EHU Av. de Tolosa 72, 20018, San Sebastián, Spain

[5]Ikerbasque Basque Foundation for Science Bilbao 481013, Spain

[6]Lead contact

[7]These authors contributed equally

\* Correspondence: derya.baran@kaust.edu.sa

\*\* Correspondence: julien.gorenflot@kaust.edu.sa

## Molecular structures

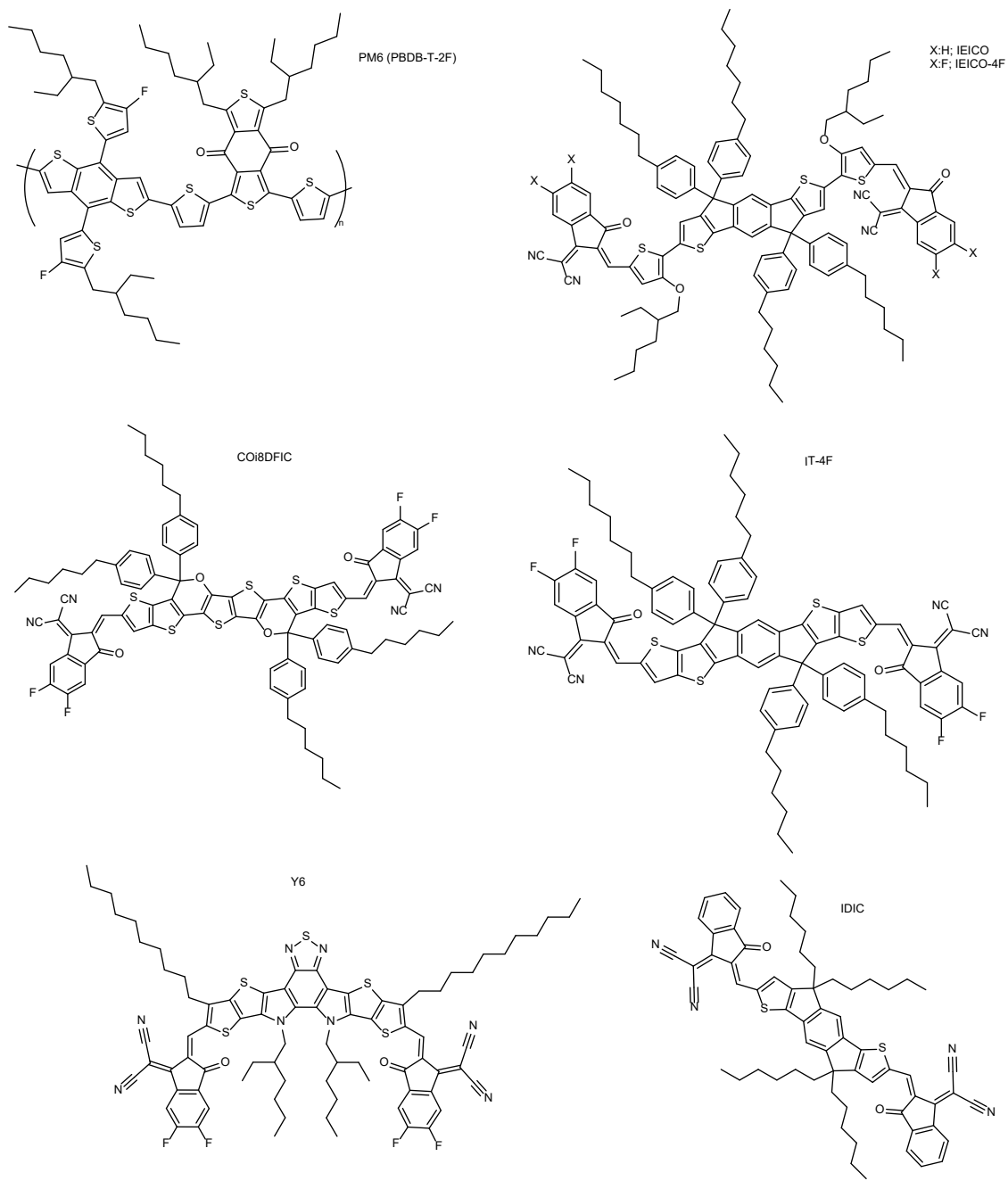

*Scheme S 1: Chemical structures of the studied molecules.*

## Absorption spectra

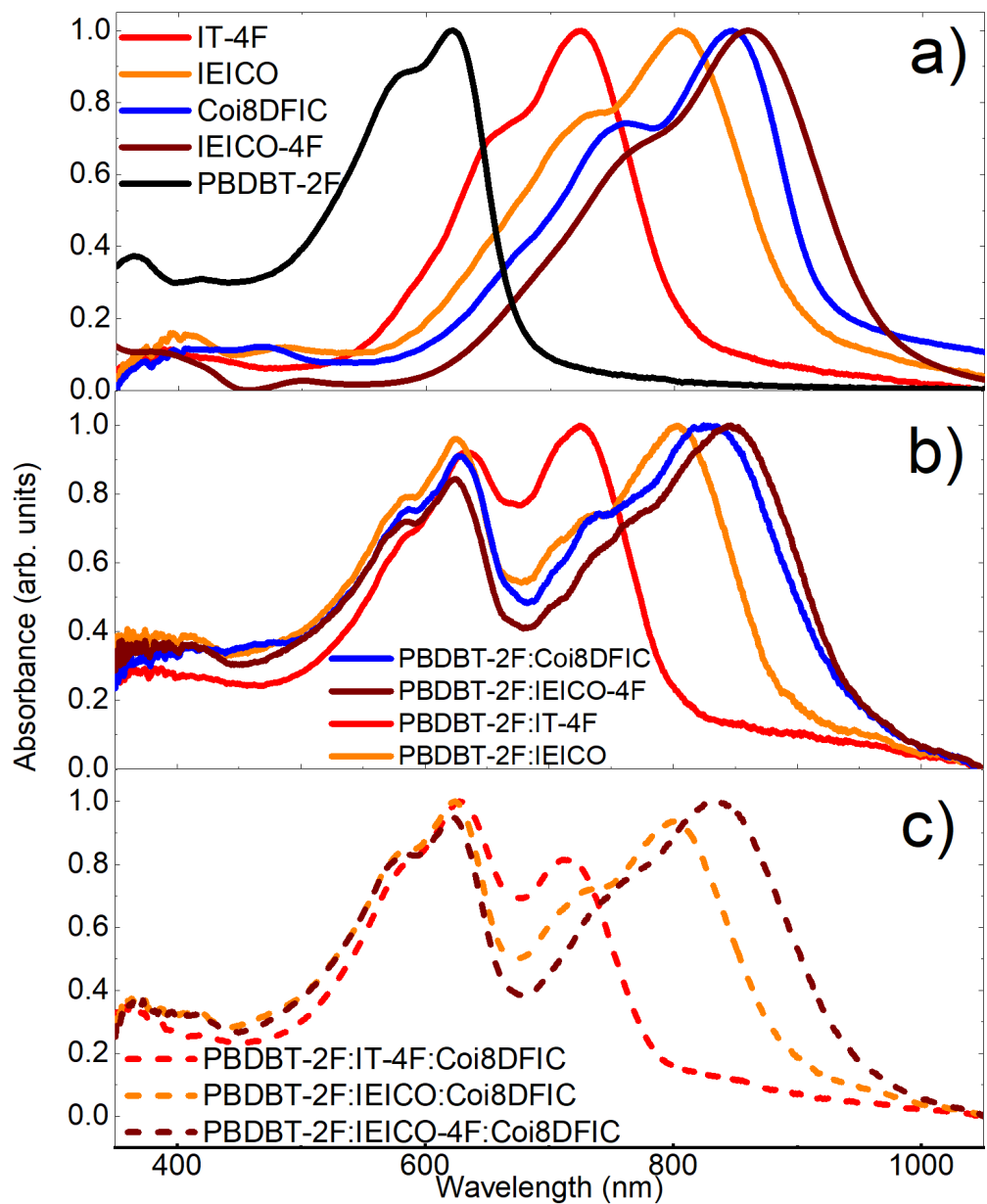

*Figure S 1: Absorption spectra of a) pristine, b) binary (1:1 ratio) and c) ternary (1:0.9:0.1 ratio) blend films.*

## UPS and LE-IPES

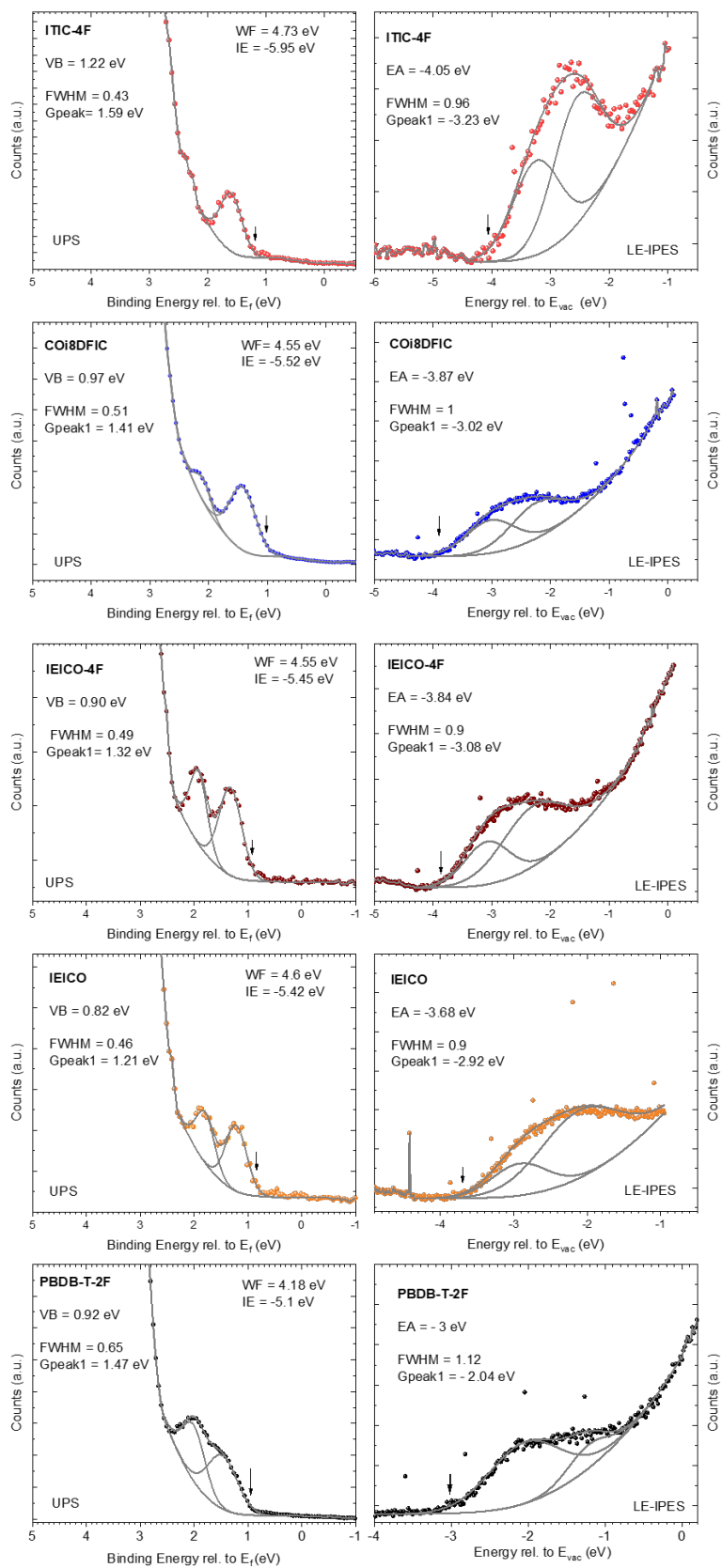

*Figure S 2: UPS and LE-IPES spectra obtained on thin films of the pristine molecules, demonstrating the onset energies of occupied (UPS) and unoccupied (LE-IPES) frontal molecular orbitals.*

Table S 1: Summary of the Ionization energy (IE) and electron affinity (EA) values measured using UPS and LE-IPES.

| Material | Ionization Potential (±0.05 eV) | Electron Affinity (±0.1 eV) |
| --- | --- | --- |
| PBDB-T-2F | 5.1 | 3.05 |
| COi8DFIC[a] | 5.6 | 3.87 |
| IT-4F | 5.9 | 4.1 |
| IEICO[a] | 5.4 | 3.75 |
| IEICO-4F[a] | 5.45 | 3.9 |

a- Those measurements were obtained by averaging the results of 2-3 UPS and LE-IPES spectra, including the one presented in Figure S 2.

## Solar cell figures of merit and statistics for the PM6:NFA-1:COi8DFIC blends

Table S 2: Device figures-of-merit obtained from JV-curves. Values are averaged over five solar cells.

| Device | Jsc (mA/cm$^2$) | Voc (V) | FF (%) | PCE (%) |
|---|---|---|---|---|
| **PM6:COi8DFIC** (1:1, B0) | 14.5±0.4 | 0.79±0.05 | 61.6±1 | 7.1±0.3 |
| **PM6:IT-4F** (1:1, B1) | 20.6±0.2 | 0.84±0.02 | 70.3±1.1 | 12.0±0.2 |
| **PM6:IT-4F:COi8DFIC** (1:0.1:0.9) | 12.9±0.5 | 0.81±0.04 | 45.7±0.2 | 4.8±0.5 |
| **PM6:IT-4F:COi8DFIC** (1:1:1) | 13.1±0.2 | 0.83±0.0 | 58.0±0.4 | 6.3±0.4 |
| **PM6:IT-4F:COi8DFIC** (1:0.9:0.1, T1) | 21.7±0.2 | 0.83±0.02 | 74.0±0.8 | 13.3±0.2 |
| **PM6:IT-4F:COi8DFIC** (1:0.5:0.5) | 20.8±0.8 | 0.84±0.01 | 57.5±0.0 | 10.1±0.4 |
| **PM6:IEICO** (1:1, B2) | 1.3±0.03 | 0.94±0.03 | 32±2 | 0.51±0.03 |
| **PM6:IEICO:COi8DFIC** (1:0.1:0.9) | 10.2±0.4 | 0.79±0.04 | 40.8±0.0 | 3.3±0.4 |
| **PM6:IEICO:COi8DFIC** (1:1:1) | 5.5±0.1 | 0.83±0.02 | 35.1±0.0 | 1.6±0.1 |
| **PM6:IEICO:COi8DFIC** (1:0.9:0.1, T2) | 1.9±0.08 | 0.94±0.01 | 31.3±1 | 0.54±0.03 |
| **PM6:IEICO:COi8DFIC** (1:0.5:0.5) | 5.0±0.4 | 0.88±0.07 | 33.3±0.0 | 1.47±0.1 |
| **PM6:IEICO-4F** (1:1, B3) | 7±0.01 | 0.84±0.01 | 49.0±1 | 2.8±0.02 |
| **PM6:IEICO-4F:COi8DFIC** (1:0.1:0.9) | 11.6±0.39 | 0.80±0.0 | 42.3±0.0 | 3.9±0.18 |
| **PM6:IEICO-4F:COi8DFIC**(1:1:1) | 5.2±0.25 | 0.72±0.05 | 40.9±3 | 1.6±0.29 |
| **PM6:IEICO-4F:COi8DFIC** (1:0.9:0.1, T3) | 8.5±0.23 | 0.84±0.01 | 49.0±1 | 3.45±0.07 |
| **PM6:IEICO-4F:COi8DFIC** (1:0.5:0.5) | 12.5±1.43 | 0.81±0.04 | 47.0±0.1 | 4.74±0.64 |

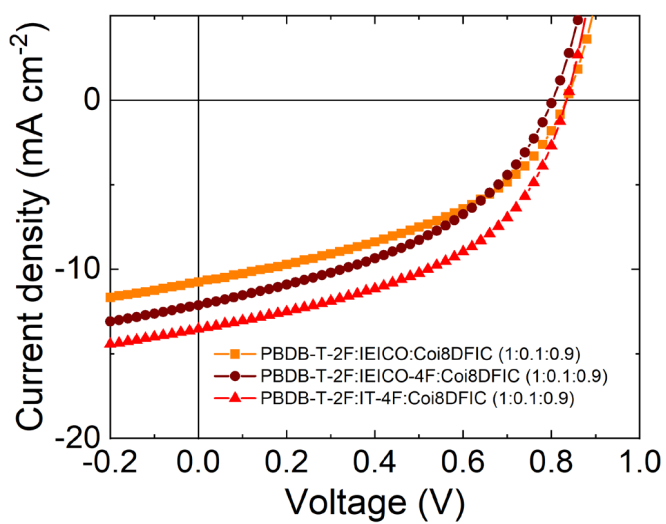

*Figure S 3: Current – Voltage characteristics of the ternary blends with blend ratio (wt.%) 1:0.1:0.9 of PM6:NFA:COi8DFIC.*

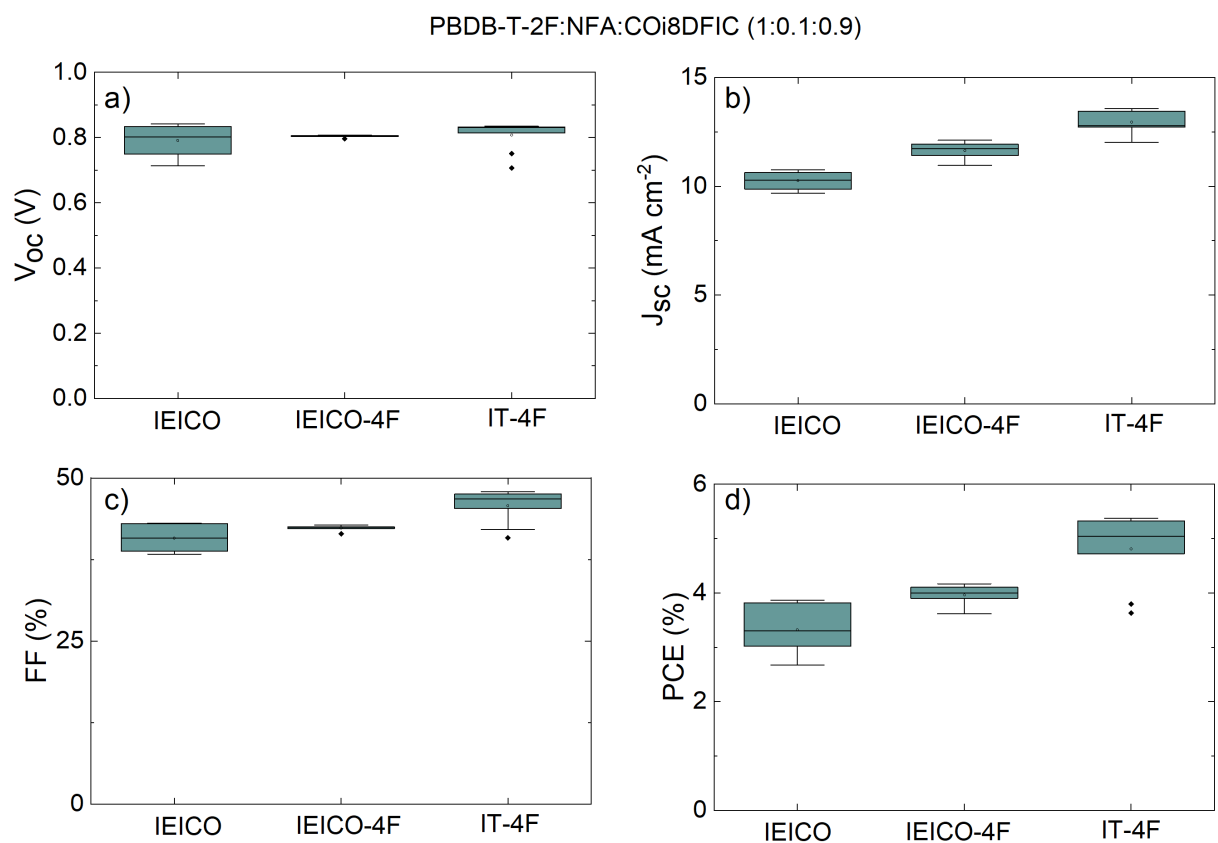

*Figure S 4: Photovoltaic parameters of different ternaries with blend ratios (wt.%) 1:0.1:0.9 of PM6:NFA:COi8DFIC.*

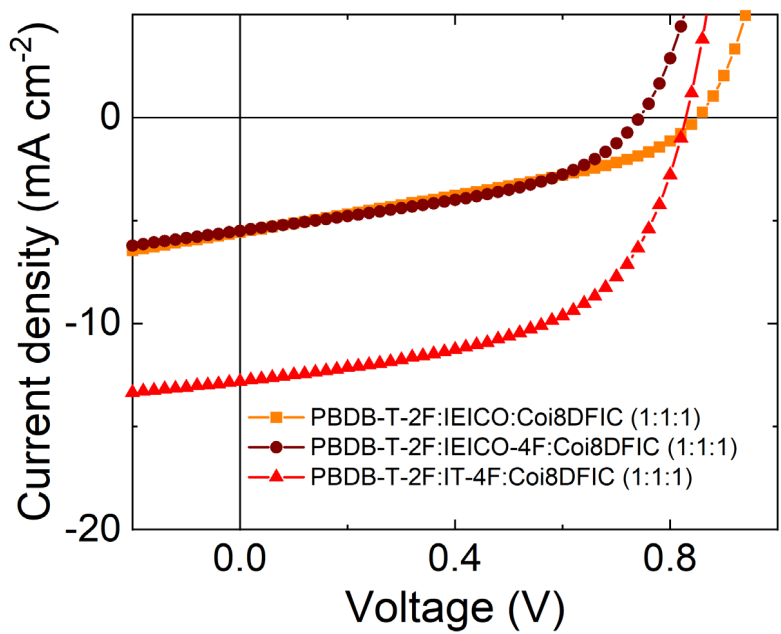

Figure S 5: Current – Voltage characteristics of the ternary blends with blend ratio (wt.%) 1:1:1 of PM6:NFA:COi8DFIC.

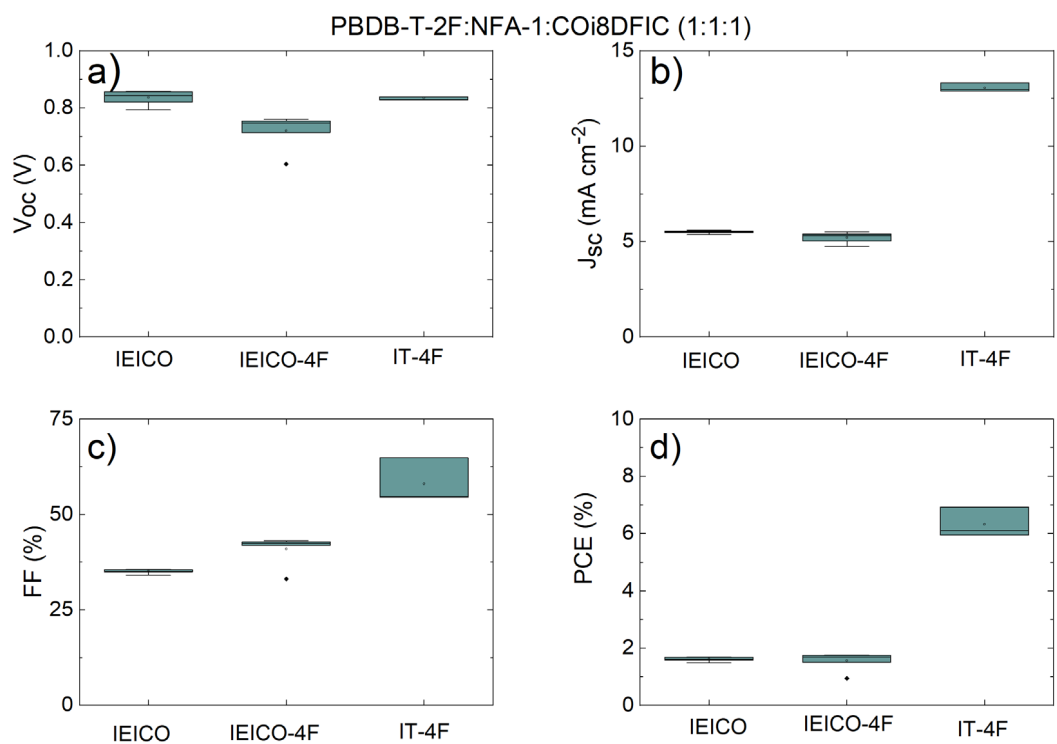

Figure S 6: Photovoltaic parameters of different ternaries with blend ratios (wt.%) 1:1:1 of PM6:NFA:COi8DFIC.

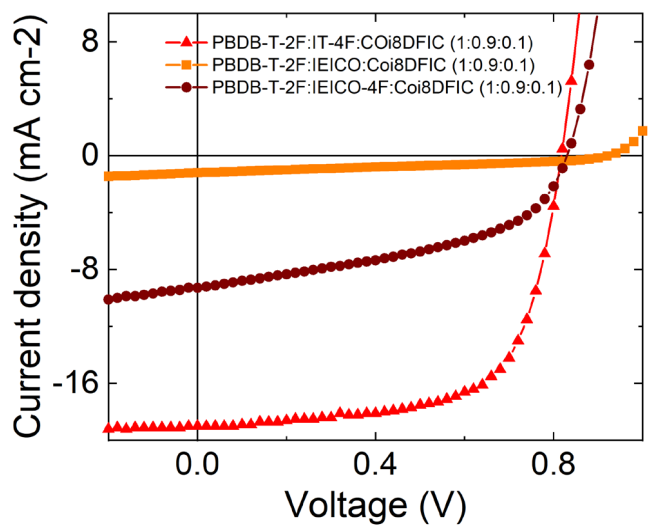

*Figure S 7: Current – Voltage characteristics of the ternary blends with blend ratio (wt.%) of 1:0.9:0.1 of PM6:NFA:COi8DFIC.*

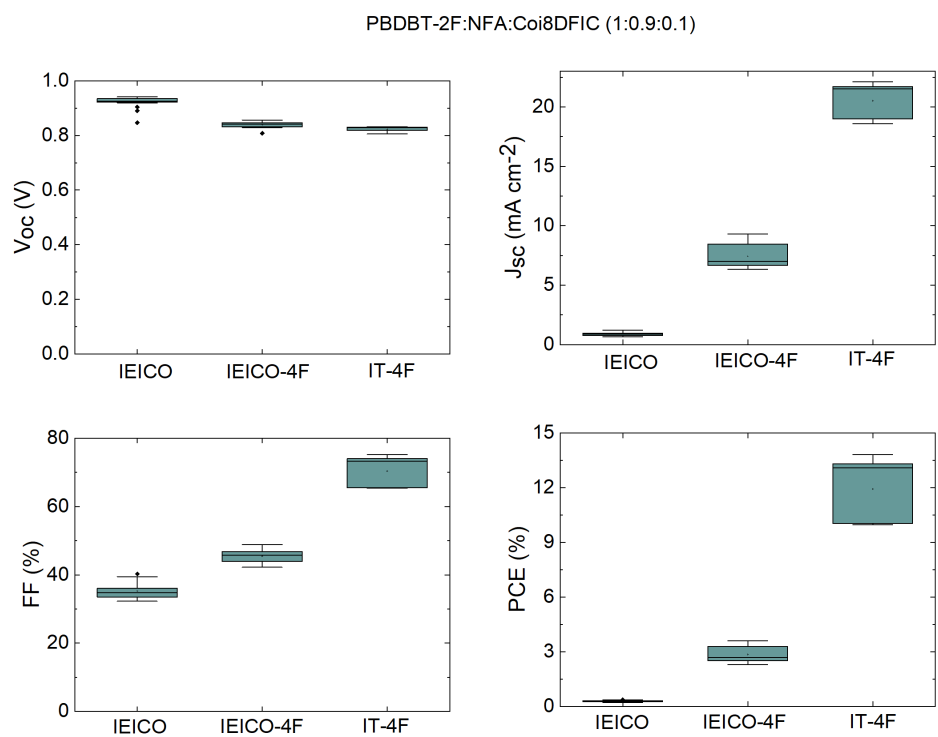

*Figure S 8: Photovoltaic parameters of different ternaries with blend ratios (wt.%) of 1:0.9:0.1 of PM6:NFA:COi8DFIC.*

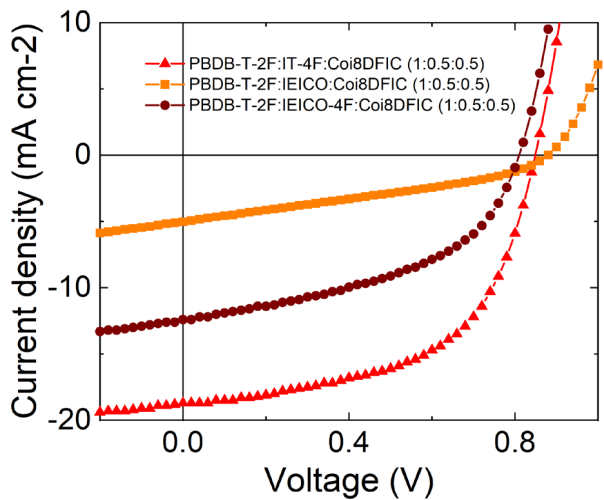

*Figure S 9: Current – Voltage characteristics of the ternary blends with blend ratio (wt.%) of 1:0.5:0.5 of PM6:NFA:COi8DFIC.*

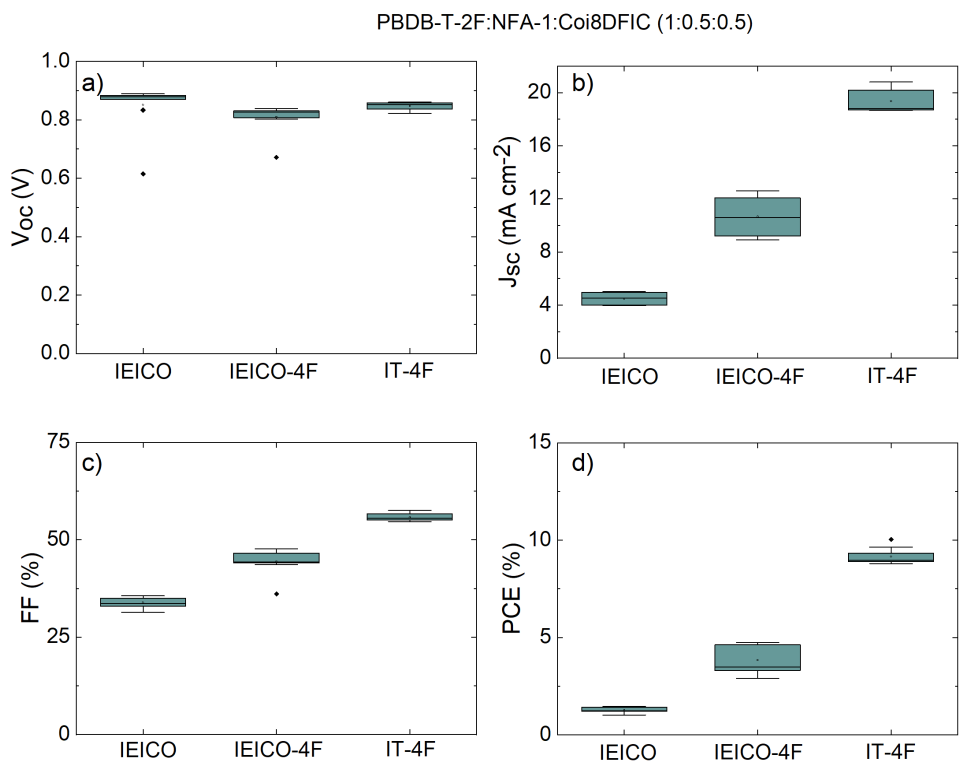

*Figure S 10: Photovoltaic parameters of different ternaries with blend ratios (wt.%) of 1:0.5:0.5 of PM6:NFA:COi8DFIC.*

IQE spectra PM6:NFA1:COi8DFIC

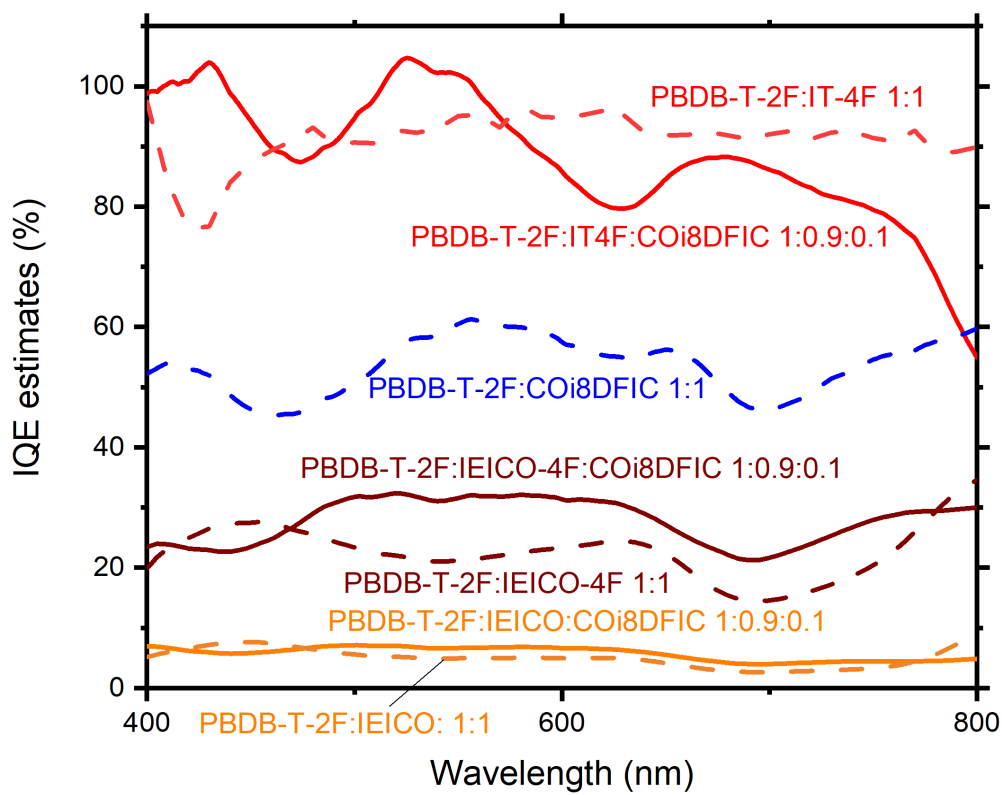

Figure S11: Estimates of the Internal quantum efficiency (IQE) spectra of PM6:NFA1:COi8DFIC systems. The values over 100% for PM6:IT-4F are most likely due to a wrong

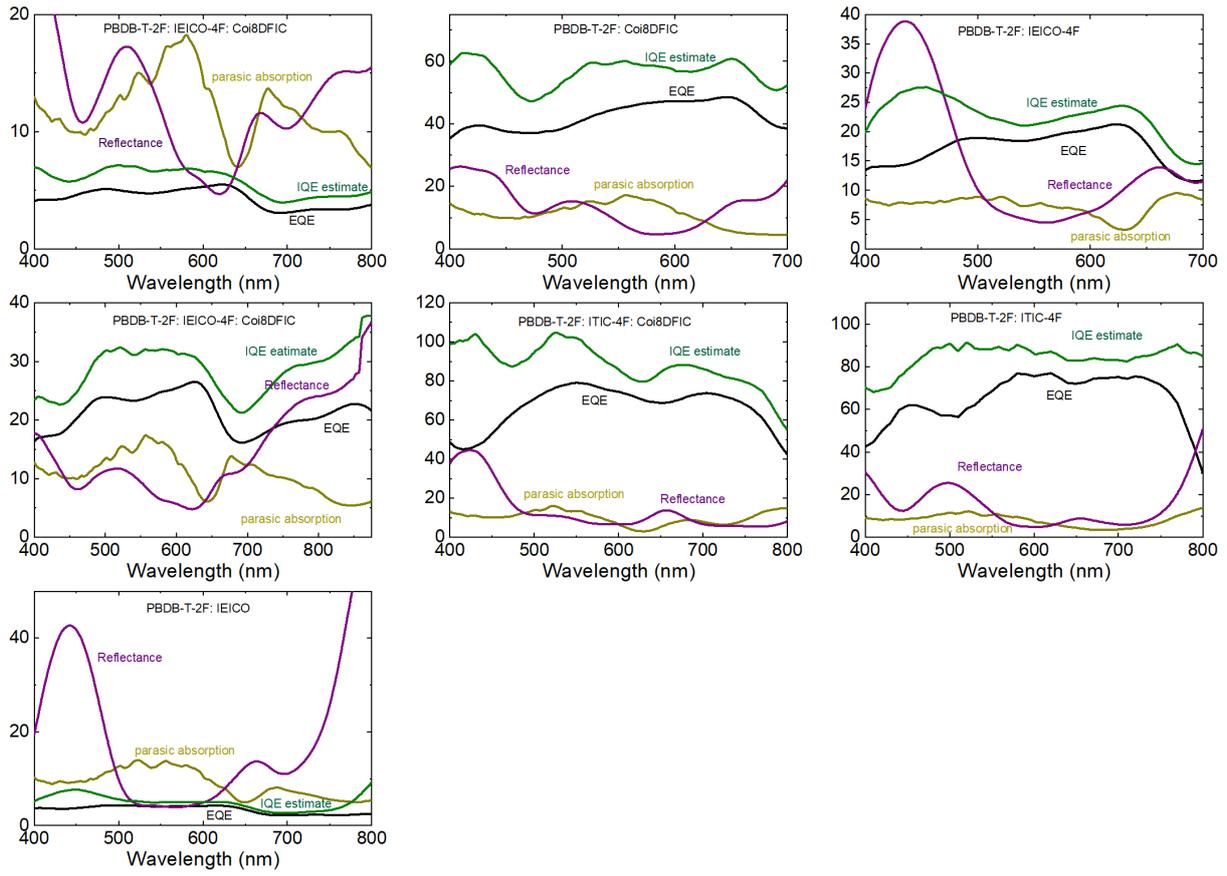

*Figure S12: Construction of the IQE$_{estimate}$ spectra for PM6:NFA-1:COi8DFIC (1:0.9:0.1) solar cells and the PM6:NFA-1 and PM6:COi8DFIC binary solar cells.*

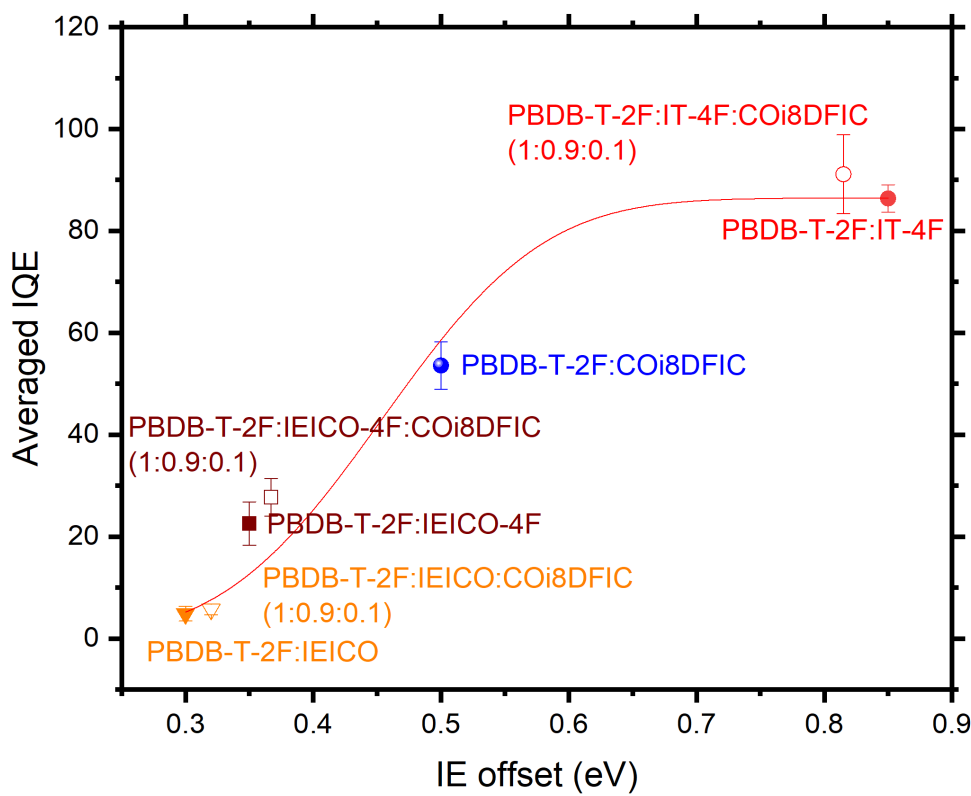

*Figure S 13: IQE vs weighted average IE offset [wt%(NFA-1) x IE(NFA-1) + wt%(COi8DFIC) x IE(COi8DFIC)]. The IQE can be well fitted (solid line) by the fraction of the interface which exhibits energy level bending lower than ΔIE, assuming a Gaussian distribution of energy level bending with an average B and a standard deviation σ, we obtain B = 0.45 ± 0.03 eV, σ =0.14 ± 0.03 eV and IQEsat = 86.4 ± 5.8 %. The error bars represent the standard deviation IQE$_{estimate}$ through the spectra (See Figure S 11).*

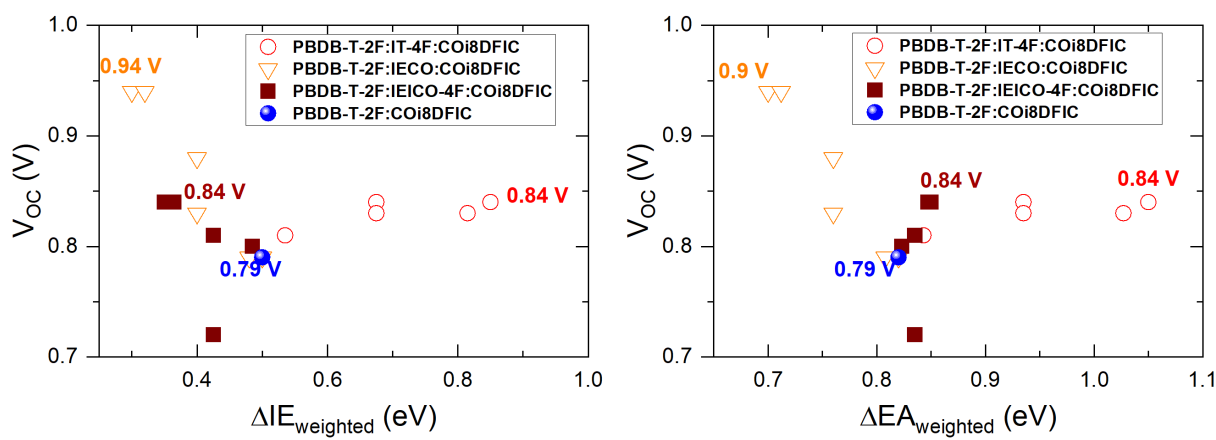

*Figure S 14: evolution of $V_{OC}$ with the weighted averaged IE and EA for the PBDB-T-2F:NFA-1:COi8DFIC blends.*

## Grazing incidence wide-angle X-ray scattering

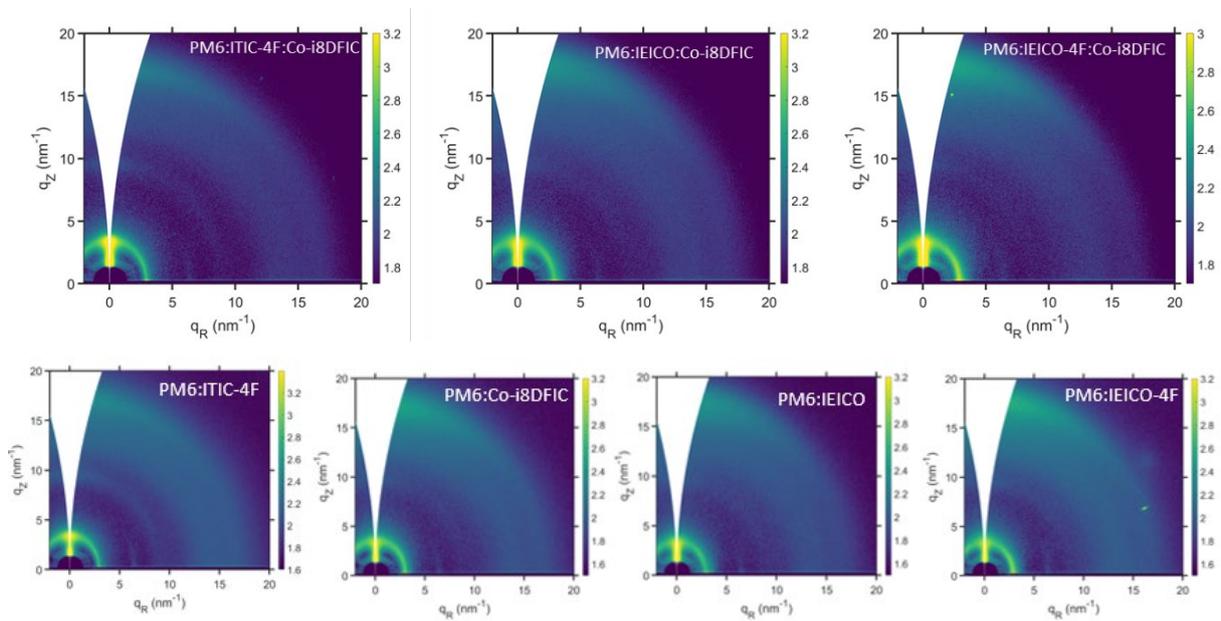

*Figure S 15: The diffractograms of the three ternaries with similar D:A1:A2 (1:0.9:0.1) ratio and respective binaries (1:1).*

## PL Spectra

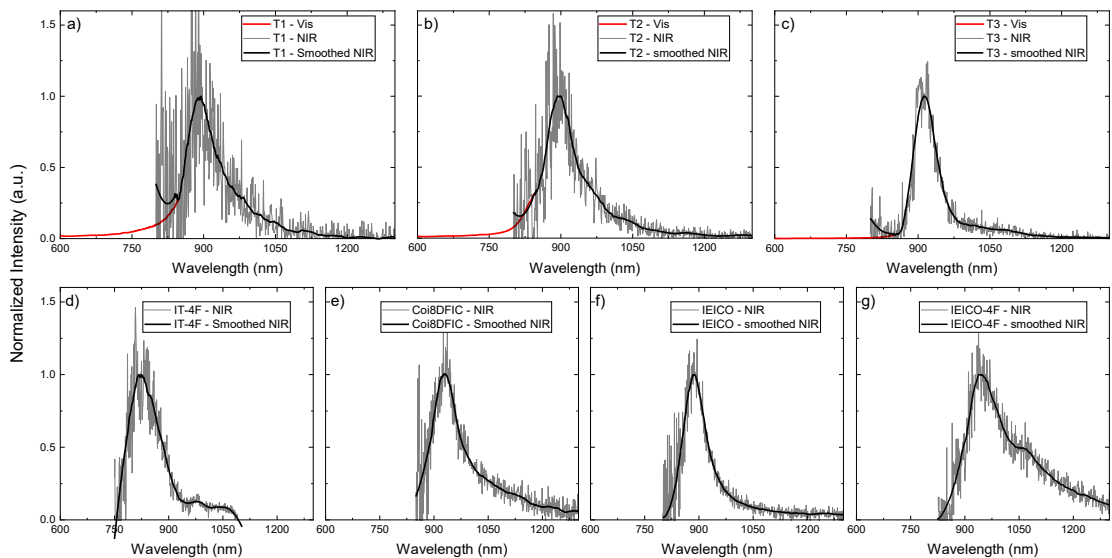

*Figure S 16. Normalized photoluminescence spectra for ternaries ($\lambda_{exc}$= 550 nm) a) T1, b) T2, and, c) T3, and neats d) IT-4F ($\lambda_{exc}$= 700 nm), Coi8DFIC ($\lambda_{exc}$= 750 nm), IEICO ($\lambda_{exc}$= 750 nm), and IEICO-4F ($\lambda_{exc}$= 790 nm). Vis and NIR in legend stand for measured with visible and NIR detector, respectively. All NIR spectra were smoothed equally for comparison (77 points Savitzky-Golay), labelled as "smoothed IR".*

## Additional transient absorption spectra and kinetics

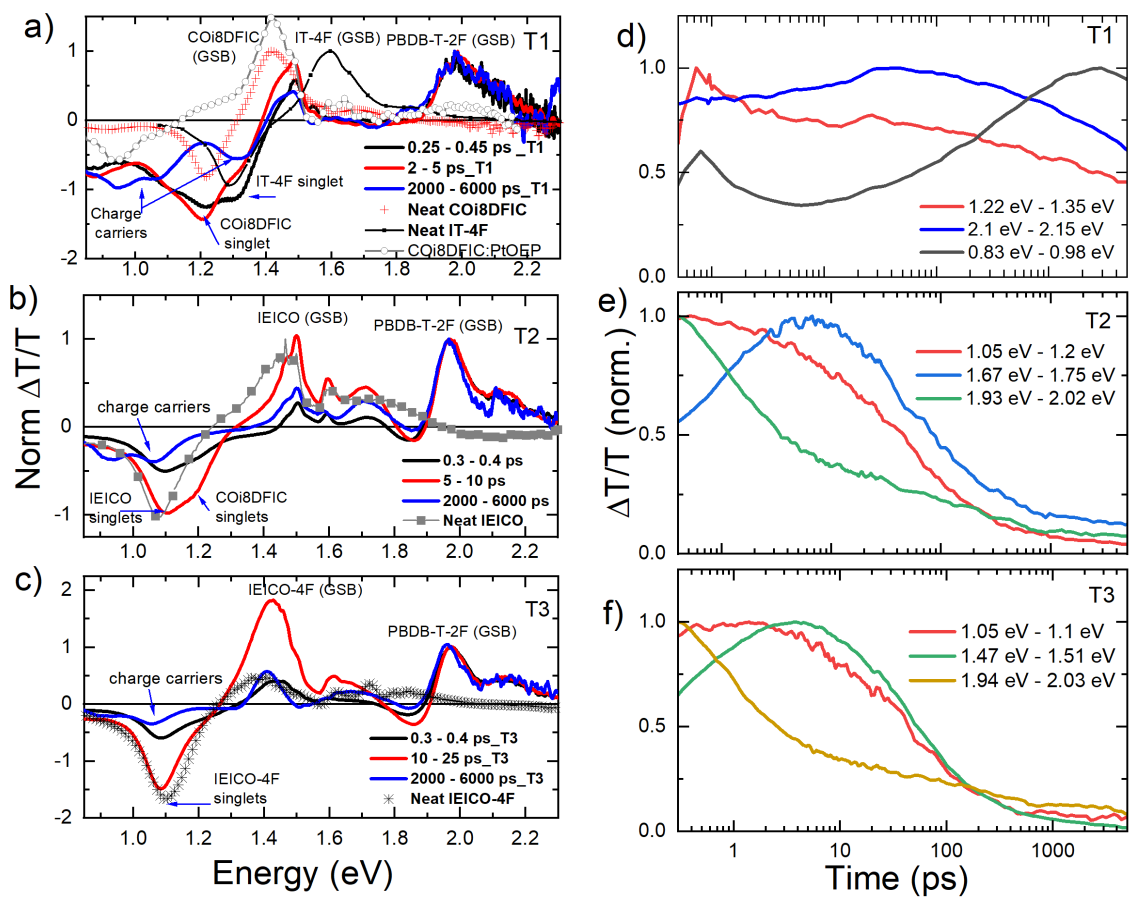

*Figure S 17: ps-ns TA spectra and kinetics after exciting at 505 nm (PM6 excitation) of T1 (a and d), T2 (b and e) and T3 (c and f).*

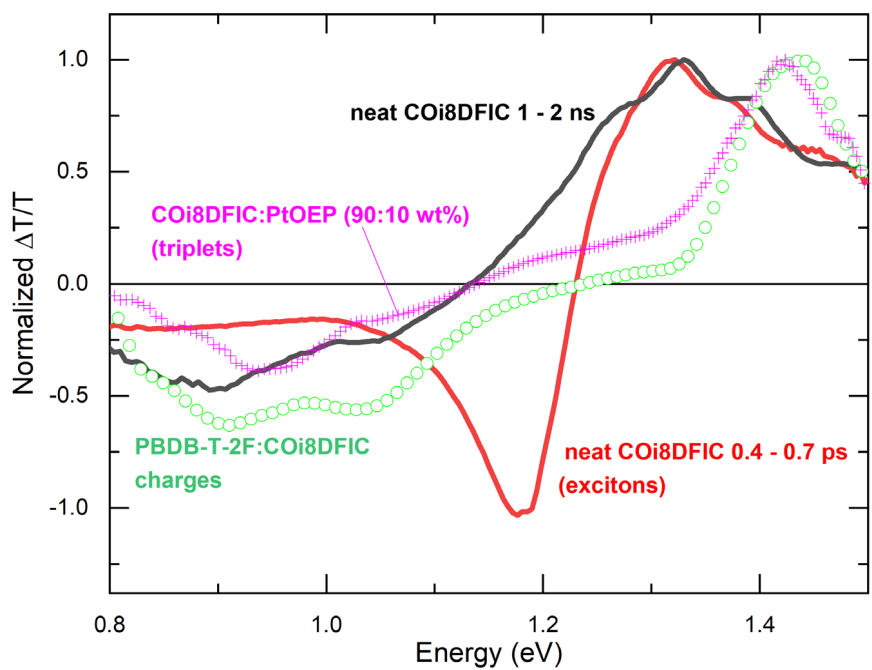

*Figure S 18: TA spectra of neat COi8DFIC after the excitation at 750 nm (red and black lines). Pink symbols show the triplet sensitized TA spectrum with Platinum octaethylporphyrin (PtOEP): COi8DFIC films with 10:90 wt%. Green open symbols represent the charge specta.*

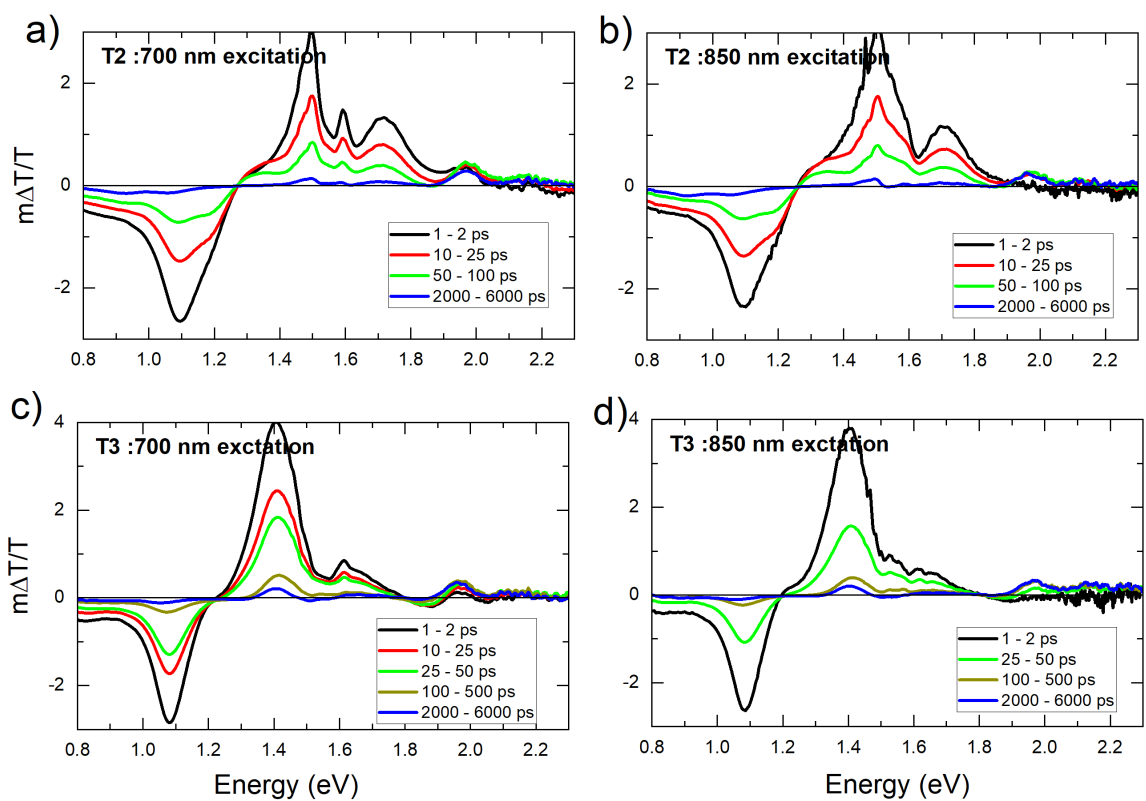

*Figure S 19: ps-ns TA spectra of T2, and T3 blend films after primarily exciting at NFA-1 at 700 nm (left column; a,c) and COi8DFIC at 850 nm (right column; b,d). Note: the large reduction of the NFA PB signal (1.4-1.6 eV) in panels a and b, is not due to a loss of excitation but to PM6 charges having a strong PA signal in the same spectral region (see Figure 3a).*

## Solar cell figures of merit for PM6:IT-4F:IEICO blends

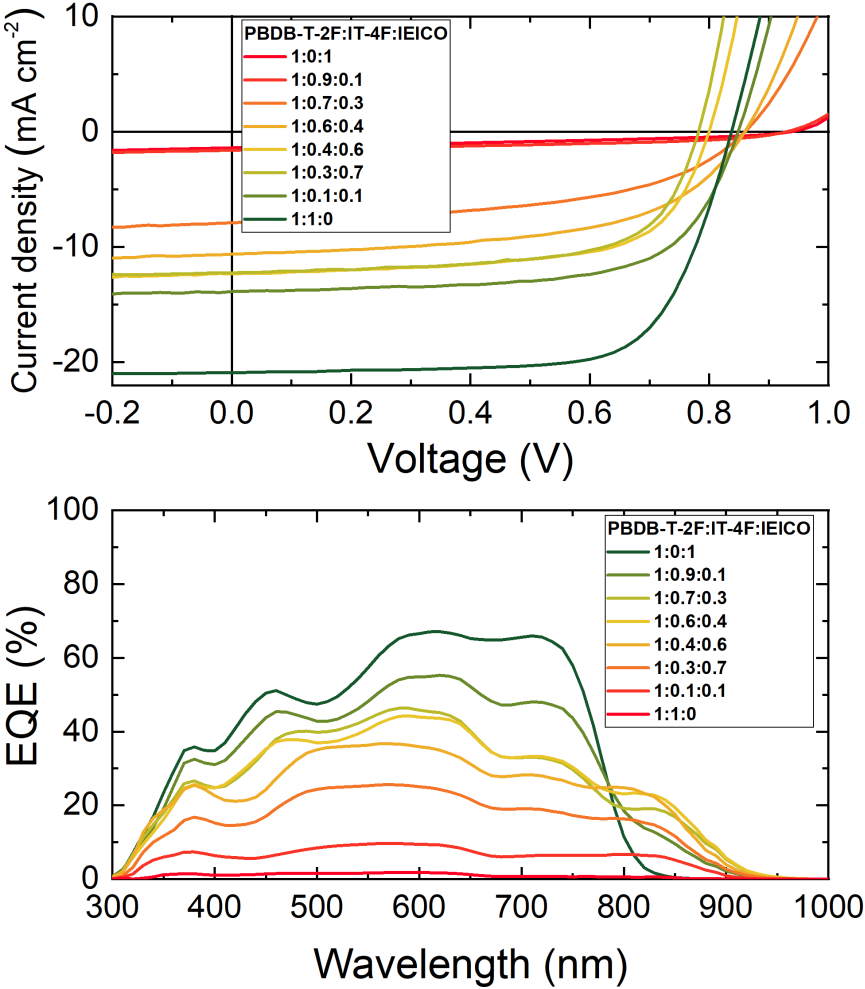

*Figure S 20: JV curves and EQE spectra of the PM6:IT-4F:IEICO ternary blends with different ratios.*

## Additional blends solar cells characteristics: PM6:Y6:IEICO and PM6:IEICO:IDIC

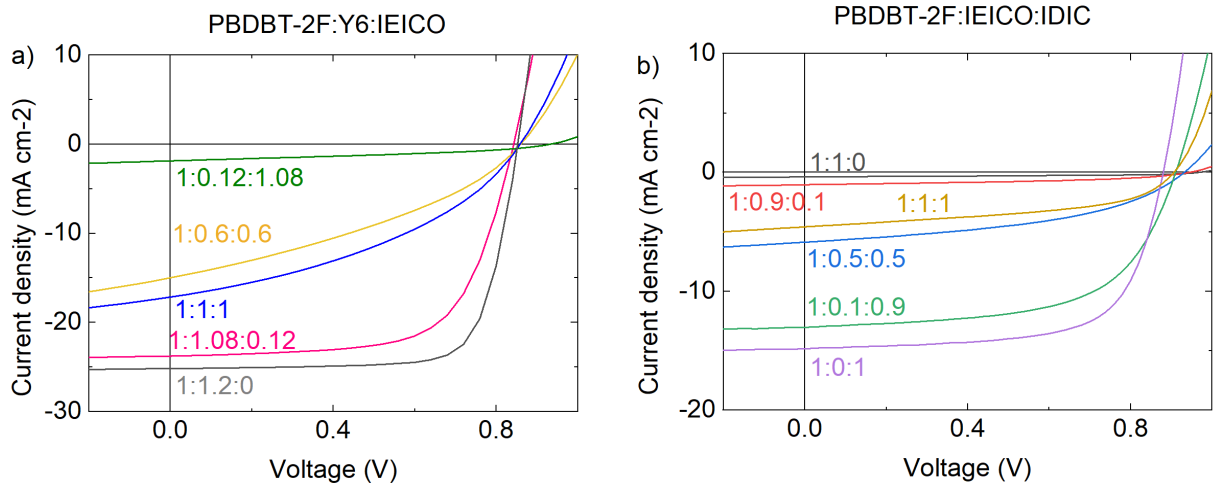

*Figure S 21: Current - voltage characteristic of the a) PM6:Y6:IEICO and b) PM6:IEICO:IDIC solar cells with different blend ratio*

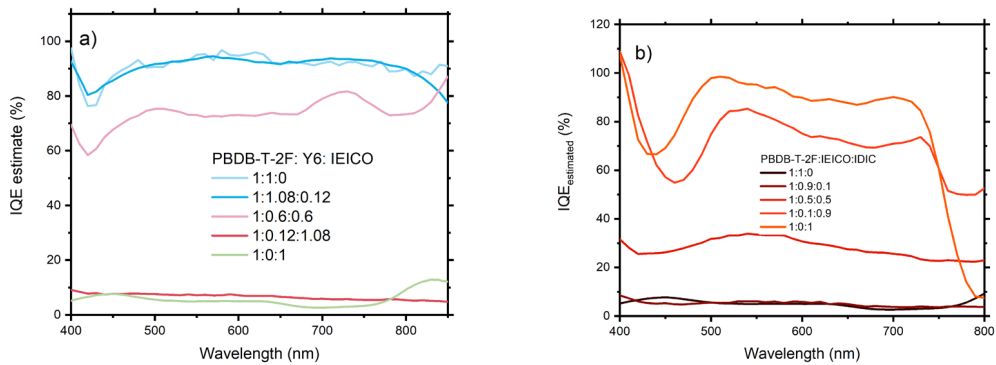

*Figure S 22: IQE$_{estimate}$ spectra of ternary PM6:Y6:IEICO (a) and PM6:IEICO:IDIC (b) systems*

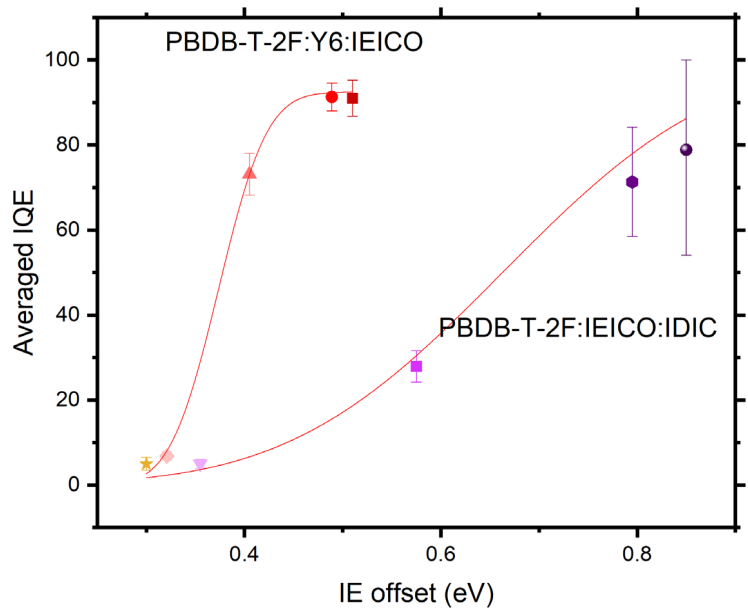

*Figure S 23: IQE vs weighted average IE offset of PM6:Y6:IEICO (a) and PM6:IEICO:IDIC (b) ternary systems. The IQE can be well fitted (solid line) by the fraction of the interface which exhibits energy level bending lower than ΔIE, assuming a Gaussian distribution of energy level bending with an average <B> and a standard deviation σ, the fitting parameters of PM6: Y6: IEICO system are B = 0.374± 0.003 eV, σ = 0.055 ± 0.004 eV and IQE$_{sat}$ = 92.5 ± 1.5 %, and B = 0.66± 0.06 eV, σ = 0.24 ± 0.07 eV and IQEsat = 100 ± 19% for PM6:IEICO:IDIC ternary system.*

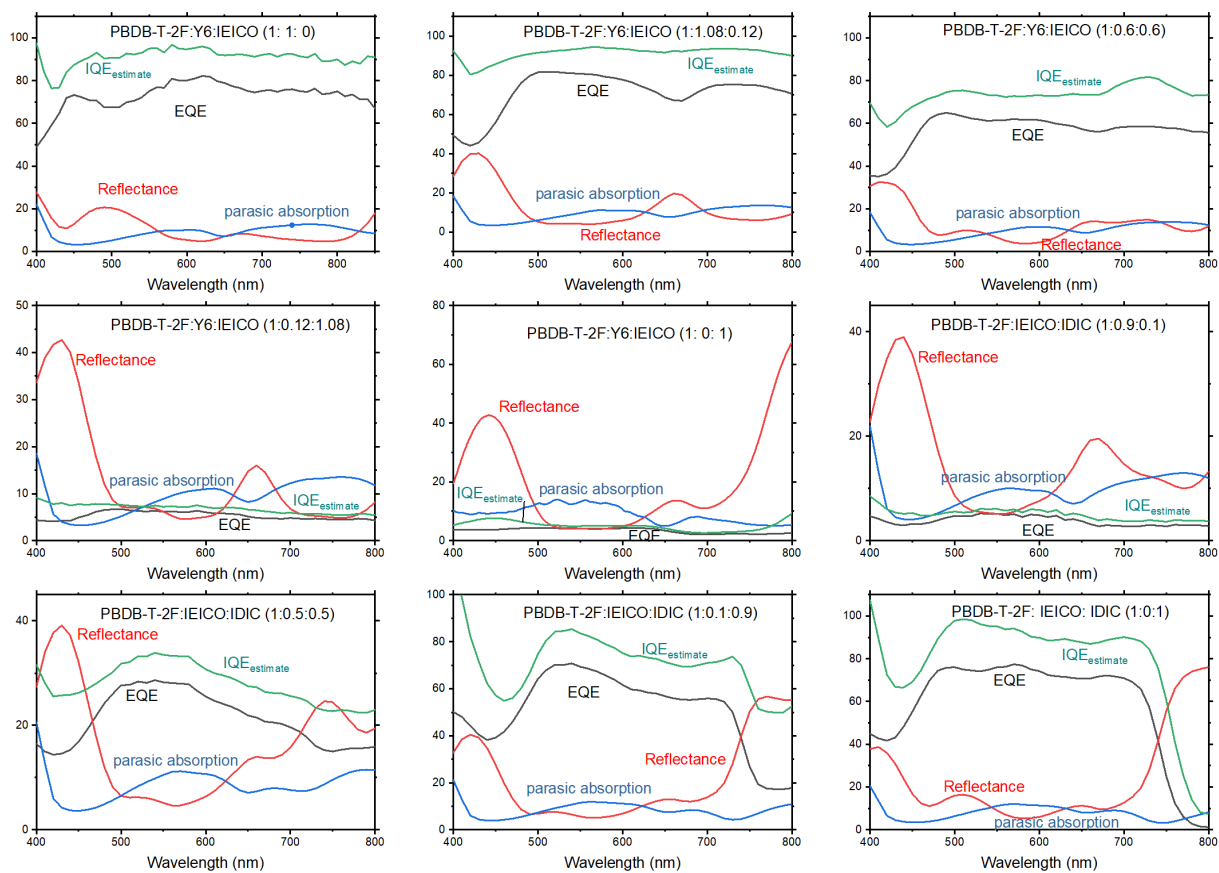

*Figure S 24: Construction of the IQE$_{estimate}$ spectra for PBDB-T-2F:6:Y6:IEICO and PBDB-T-2F::IEICO:IDIC based ternary and binary solar cells.*

*Table S3:: Device figures-of-merit obtained from JV-curves of PM6:Y6:IEICO devices. Values are averaged over five solar cells.*

| Blend | $J_{sc}$ (mA cm$^{-2}$) | $V_{oc}$ (mV) | FF (%) | PCE (%) |
| --- | --- | --- | --- | --- |
| **PM6:Y6 (1:1.2:0)** | 24.9 ± 0.2 | 850 ± 0.01 | 74.2 ± 0.02 | 15.8 ± 0.5 |
| **PM6:Y6:IEICO (1:1.08:0.12)** | 23.8 ± 0.1 | 842 ± 2 | 65.5 ± 0.01 | 13.1 ± 0.07 |
| **PM6:Y6:IEICO (1:0.6:0.6)** | 14.9 ± 0.3 | 859 ± 4 | 35.1 ± 0.01 | 4.5 ± 0.15 |
| **PM6:Y6:IEICO (1:1:1)** | 16.9 ± 0.4 | 858 ± 3 | 38.2 ± 0.01 | 5.54 ± 0.24 |
| **PM6:Y6:IEICO (1:0.12:1.08)** | 1.9 ± 0.7 | 938 ± 3 | 37.01 ± 0.01 | 0.67 ± 0.01 |

Table S4: : Device figures-of-merit obtained from JV-curves of PM6:IEICO:IDIC devices. Values are averaged over five solar cells.

| Blend | $J_{sc}$ (mA cm$^{-2}$) | $V_{oc}$ (mV) | FF (%) | PCE (%) |
|---|---|---|---|---|
| **PM6:IEICO (1:1:0)** | 0.4 ± 0.01 | 962 ± 3 | 46.6 ± 0.02 | 0.2 ± 0.06 |
| **PM6:IEICO:IDIC (1:0.1:0.9)** | 13.0 ± 0.1 | 909 ± 3 | 59.6 ± 0.01 | 7.1 ± 0.2 |
| **PM6:IEICO:IDIC (1:0.5:0.5)** | 5.7 ± 0.1 | 932 ± 2 | 44.3 ± 0.03 | 2.4 ± 0.1 |
| **PM6:IEICO:IDIC (1:1:1)** | 4.6 ± 0.07 | 906 ± 10 | 48.2 ± 0.03 | 2.0 ± 0.03 |
| **PM6:IEICO:IDIC (1:0.9:0.1)** | 1.1 ± 0.03 | 869 ± 10 | 40.3 ± 0.06 | 0.38 ± 0.12 |
| **PM6:(IEICO):IDIC (1:0:1)** | 14.8 ± 0.02 | 875 ± 4 | 66.6 ± 0.01 | 8.6 ± 0.1 |